\documentclass[draftcls,twoside,onecolumn,letter,12pt]{IEEEtran}
\usepackage{graphicx}
\usepackage{amssymb}
\usepackage{amsmath}
\usepackage{cite}
\usepackage{cases}

\newtheorem{theorem}{Theorem}
\newtheorem{lemma}{Lemma}
\newtheorem{corollary}{Corollary}
\newtheorem{proposition}{Proposition}

\DeclareMathOperator{\vecOp}{\mathrm{vec}}
\DeclareMathOperator{\tr}{\mathrm{tr}}
\DeclareMathOperator{\E}{\mathbb{E}}

\DeclareMathOperator{\cov}{\mathbb{C}\mathrm{ov}}

\DeclareMathOperator{\etr}{\mathrm{etr}}

\newcommand{\EX}[1]{\E\left\{{#1}\right\}}

\newcommand{\EXs}[2]{\E_{{#1}}\left\{{#2}\right\}}
\newcommand{\PDF}[2]{p_{{#1}}\left({#2}\right)}

\newcommand{\B}[1]{\pmb{#1}}
\newcommand{\C}{\mathbb{C}}
\newcommand{\R}{\mathbb{R}}

\newcommand{\CGM}[5]{\tilde{\mathcal{N}}_{{#1},{#2}}\left({#3},{#4},{#5}\right)}

\newcommand{\CQM}[5]{\tilde{\mathcal{Q}}_{{#1},{#2}}\left({#3},{#4},{#5}\right)}

\newcommand{\nTx}{{n_{\mathrm{T}}}}
\newcommand{\nRx}{{n_{\mathrm{R}}}}
\newcommand{\TxCM}{\B{\Phi}_{\mathrm{T}}}
\newcommand{\RxCM}{\B{\Phi}_{\mathrm{R}}}
\newcommand{\TxCC}{\zeta_{\mathrm{T}}}
\newcommand{\RxCC}{\zeta_{\mathrm{R}}}
\newcommand{\Tc}{{N_{\text{c}}}}
\newcommand{\Nb}{{N_{\text{b}}}}
\newcommand{\nXmin}{m}
\newcommand{\nXmax}{n}

\newcommand{\CMmin}{\B{\Phi}_{1}}
\newcommand{\CMmax}{\B{\Phi}_{2}}

\newcommand{\MHH}{\B{\Theta}}
\newcommand{\Er}{{E_\text{r}}}
\newcommand{\Eg}{E_{0,\mathcal{\tilde{N}}}}
\newcommand{\Rcr}{R_{\text{cr}}}
\newcommand{\EoRF}[4]{\mathcal{L}_0^{#4}\left({#1},{#2},{#3}\right)}
\newcommand{\EoRFA}[4]{\mathcal{L}_1^{#4}\left({#1},{#2},{#3}\right)}
\newcommand{\OB}[1]{\beta^\ast\left(#1\right)}
\newcommand{\ErC}{\left\langle C \right\rangle}

\newcommand{\snr}{\gamma}

\newcommand{\HyperPFQ}[3]{{}_{#1}F_{#2}\left(#3\right)}
\newcommand{\PS}[2]{\left(#1\right)_{{#2}}}
\newcommand{\matHyperPFQII}[4]{{}_{#1}\tilde{F}_{#2}^{\left(#3\right)}\left(#4\right)}

\newcommand{\EoT}[3]{\tilde{E}_0\left(#1,#2,#3\right)}
\newcommand{\EoConst}[2]{\mathcal{K}\left(#1,#2\right)}
\newcommand{\dBEoConst}[2]{\mathcal{K}^{\left(\beta\right)}\left(#1,#2\right)}
\newcommand{\dREoConst}[2]{\mathcal{K}^{\left(\rho\right)}\left(#1,#2\right)}
\newcommand{\Kc}{K_\text{cor}}
\newcommand{\Kiid}{K_\text{iid}}

\newcommand{\EVn}[1]{{\varrho\left(#1\right)}}
\newcommand{\EVm}[2]{{\chi_{#1}\left(#2\right)}}

\newcommand{\EVV}[1]{{\pmb{\lambda}\left(#1\right)}}
\newcommand{\EV}[3]{{\lambda_{#1}^{#2}\left(#3\right)}}
\newcommand{\oEV}[3]{{\lambda_{\langle #1 \rangle}^{#2}\left(#3\right)}}

\newcommand{\rbSnr}{\eta}

\newcommand{\mySepR}{\vspace{-0.1cm}}

\newcommand{\mySep}{\vspace*{2pt}}
\newcommand{\EqF}{}
\newcommand{\EqE}{}

\newcommand{\PFC}[3]{\mathcal{X}_{#1,#2}\left(#3\right)}
\newcommand{\CM}[2]{\B{\mathsf{G}}_{(#1)}\left(#2\right)}

\date{December, 2005}

\makeatletter
\def\@setsize#1#2#3#4{
    \@nomath#1
    \let\@currsize#1
    \baselineskip #2
    \baselineskip \baselinestretch\baselineskip
    \parskip \baselinestretch\parskip
    \setbox\strutbox \hbox{
        \vrule height.7\baselineskip
            depth.3\baselineskip
            width\z@}
    \skip\footins \baselinestretch\skip\footins
    \normalbaselineskip\baselineskip#3#4}
\makeatother

\makeatletter
\newcommand{\setstretch}[1]{
    \def\baselinestretch{#1}%
    \@currsize
    }
\makeatother

\makeatletter
\newenvironment{salign}{
    \vskip 0.0\baselineskip
    \setstretch{1}
    \start@align\@ne\st@rredfalse\m@ne
    }{
    \math@cr
    \black@\totwidth@
    \egroup
    \ifingather@
        \restorealignstate@
        \egroup
        \nonumber
        \ifnum0=`{\fi\iffalse}\fi
        \else
        $$%
    \fi
    \ignorespacesafterend
    \vskip 0.6\baselineskip
    \par
    \noindent
    }
\makeatother

\makeatletter
\newenvironment{sgather}{
    \vskip 0.0\baselineskip
    \setstretch{1}
    \start@gather\st@rredfalse
    }{
    \math@cr \black@\totwidth@ \egroup
    $$\ignorespacesafterend
    \vskip 0.6\baselineskip
    \par
    \noindent
    }
\makeatother

\begin{document}
\title{
        \hspace{2cm}\\[1.5cm]
        Gallager's Exponent for MIMO Channels:\\ A Reliability--Rate
        Tradeoff
        }
\author{
        \hspace{4cm}\\
        Hyundong Shin, \IEEEmembership{Member, IEEE},
        and
        Moe Z. Win, \IEEEmembership{Fellow, IEEE}
\\[1.5cm]
        \underline{Corresponding Author:} \\[0.3cm]
        Hyundong Shin \\[0.1cm]
        Laboratory for Information and Decision Systems (LIDS)
        \\[-0.1cm]
        Massachusetts Institute of Technology\\[-0.1cm]
        Room 32-D782, 77 Massachusetts Avenue\\[-0.1cm]
        Cambridge, MA 02139 USA \\[0.2cm]
        Tel.: (617) 253-6173 \\[-0.1cm]
        e-mail: \tt hshin@mit.edu\\
\thanks{This research was supported, in part, by
            the Korea Research Foundation Grant funded by the
            Korean Government (KRF-2004-214-D00337),
            the Charles Stark Draper Laboratory Robust Distributed Sensor Networks Program,
            the Office of Naval Research Young Investigator Award N00014-03-1-0489,
            and
            the National Science Foundation under Grant ANI-0335256.
            }
\thanks{H. Shin and M. Win are with
            the Laboratory for Information and Decision Systems (LIDS),
            Massachusetts Institute of Technology,
            77 Massachusetts Avenue,
            Cambridge, MA 02139 USA
            (e-mail: {\tt hshin@mit.edu},
                {\tt moewin@mit.edu}).}
}

\markboth{Submitted to the IEEE Transactions on
          Communications}
         {Shin and Win:
          Gallager's Exponent for MIMO Channels: A Reliability--Rate Tradeoff}

\maketitle

\clearpage

\begin{abstract}
In this paper, we derive Gallager's random coding error exponent
for multiple-input multiple-output (MIMO) channels, assuming no
channel-state information (CSI) at the transmitter and perfect CSI
at the receiver. This measure gives insight into a fundamental
tradeoff between the \emph{communication reliability} and
\emph{information rate} of MIMO channels, enabling to determine
the required codeword length to achieve a prescribed error
probability at a given rate below the channel capacity. We
quantify the effects of the number of antennas, channel coherence
time, and spatial fading correlation on the MIMO exponent.  In
addition, general formulae for the ergodic capacity and the cutoff
rate in the presence of spatial correlation are deduced from the
exponent expressions. These formulae are applicable to arbitrary
structures of transmit and receive correlation, encompassing all
the
previously known results as special cases of our expressions. 

\end{abstract}


\section{Introduction}  \label{sec:Sec:1}


The channel capacity is a crucial information-theoretic
perspective that determines the fundamental limit on achievable
information rates over a communication channel \cite{Sha:48:BSTJ}.
However, since the channel capacity alone gives only the knowledge
of the maximum achievable rate, a stronger form of the channel
coding theorem has been pursued to determine the behavior of the
error probability $P_\mathrm{e}$ as a function of the codeword
length $N$ and information rate $R$
\cite{Sha:59:BSTJ,Gal:68:Book,Eri:70:IT}. The \emph{reliability
function} or the \emph{error exponent} of a communication system
is defined by \cite{Sha:59:BSTJ}
\begin{align*}
\EqF
    E\left(R\right)
    \triangleq
        \limsup_{N \to \infty}
            \frac{
                    -\ln
                    P_\text{e}^\text{opt}\left(R,N\right)
                 }
                 {N}
\EqE
\end{align*}
where $P_\text{e}^\text{opt}\left(R,N\right)$ is the average block
error probability for the optimal block code of length $N$ and
rate $R$.\footnote{
        In the following, we will use the term ``error probability" to denote the average block
        error probability.
        }
The error exponent describes a decaying rate in the error
probability as a function of the codeword length, and hence serves
to indicate how difficult it may be to achieve a certain level of
reliability in communication at a rate below the channel capacity.
Although it is difficult to find the exact error exponent, its
classical lower bound is available due to Gallager
\cite{Gal:68:Book}. This lower bound is known as the \emph{random
coding error exponent} or \emph{Gallager's exponent} in honor of
his discovery, and has been used to estimate the codeword length
required to achieve a prescribed error probability
\cite{Ahm:97:PhD,AM:99:IT,AH:99:TM}.

The random coding exponent was extensively studied for
single-input single-output (SISO) and single-input multiple-output
(SIMO) flat-fading channels with average or peak power constraint
\cite{Ahm:97:PhD,AM:99:IT}. For SIMO block-fading channels, the
random coding exponent was derived in \cite{KS:95:AEU} with
perfect channel-state information (CSI) at the receiver, where it
has been shown that although the capacity is independent of the
channel coherence time (first asserted in \cite{MS:84:IT} and also
recently addressed in \cite{MH:99:IT} and \cite{HMH:01:IT} for
multiple-antenna communication), the error exponent suffers a
considerable decrease due to a reduction in the effective codeword
length as the coherence time increases.\footnote{
        This observation is parallel to the \emph{divergent} behavior of the
        channel capacity and cutoff rate of a channel with block memory
        \cite{MS:84:IT}.
        }
Therefore, this so-called \emph{channel-incurable effect} reduces
the communication reliability. While there are numerous prior
investigations (following the seminal work of
\cite{Win:87:JSAC,WSG:94:COM,Fos:96:BLTJ,FG:98:WPC}) on the
capacity for multiple-input multiple-output (MIMO) channels
\cite{Tel:99:ETT,SFGK:00:COM,CTKV:02:IT,CWZ:03:IT,SRS:03:IT,SL:03:IT,SWLC:05:WCOM,SW:05:IT},
only limited results are available for error exponents. The random
coding exponent were given implicitly in \cite{Tel:99:ETT}
(without final analytical expressions) for independent and
identically distributed (i.i.d.) Rayleigh-fading MIMO channels
with a single-symbol coherence time, perfect receive CSI, and
Gaussian inputs subject to the average power constraint. Also, the
random coding exponent was analyzed in \cite{AH:99:TM} for i.i.d.\
block-fading MIMO channels with no CSI and isotropically unitary
inputs subject to the average power constraint. 

In this paper, taking into account spatial fading correlation, we
derive Gallager's exponent for MIMO channels. We consider a
block-fading channel with Gaussian inputs subject to the average
power constraint and perfect CSI at the receiver. Our results
resort to the methodology developed in \cite{SWLC:05:WCOM} and
\cite{SW:05:IT}, which is based on the finite random matrix theory
\cite{Jam:64:AMS,Kha:66:AMS}.
The MIMO exponent obtained in the paper provides insight into a
fundamental tradeoff between the communication reliability and
information rate (below the channel capacity), enabling to
determine the required codeword length for a prescribed error
probability. It is interesting to note that as a special case of
this \emph{reliability--rate} tradeoff, one can obtain the
diversity--multiplexing tradeoff of MIMO channels
\cite{Win:87:JSAC,WSG:94:COM,ZT:03:IT}, which is a scaled version
of the asymptotic reliability--rate tradeoff at
high signal-to-noise ratio (SNR). 
%
%
We quantify the effects of the number of antennas, the channel
coherence time, and the amount of spatial fading correlation on
the MIMO exponent. Moreover, the general formulae for the ergodic
capacity and cutoff rate are deduced from the exponent
expressions. In particular, our capacity formula embraces all the
previously known results for i.i.d.\ \cite{Tel:99:ETT,SL:03:IT},
one-sided correlated \cite{CWZ:03:IT,SRS:03:IT}, and doubly
correlated \cite{SWLC:05:WCOM} channels.

This paper is organized as follows. In Section~\ref{sec:Sec:2},
signal and channel models are presented. Section~\ref{sec:Sec:3}
derives the expression for the MIMO random coding exponent.
Section~\ref{sec:proof} gives proofs of the main results stated in
Theorem~\ref{thm:RCEE}. In Section~\ref{sec:Sec:4}, some numerical
results are provided to illustrate the reliability--rate tradeoff
in block-fading MIMO channels. Finally, Section~\ref{sec:Sec:5}
concludes the paper.


\textit{Notation:} Throughout the paper, we shall use the
following notation. $\mathbb{N}$, $\R$, and $\C$ denote the
natural numbers and the fields of real and complex numbers,
respectively. The superscripts $T$ and $\dag$ stand for the
transpose and transpose conjugate, respectively. $\B{I}_n$ is the
$n \times n$ identity matrix and $\left(A_{ij}\right)$ denotes the
matrix with the $\left(i,j\right)$th entry $A_{ij}$. The trace
operator of a square matrix $\B{A}$ is denoted by
$\tr\left(\B{A}\right)$ and
$\etr\left(\B{A}\right)=e^{\tr\left(\B{A}\right)}$. The Kronecker
product of matrices is denoted by $\otimes$. By $\B{A}
>0$, we denote $\B{A}$ is positive definite. For a Hermitian matrix $\B{A} \in
\C^{n \times n}$, $\EV{1}{}{\B{A}} \geq \EV{2}{}{\B{A}} \geq
\ldots \geq \EV{n}{}{\B{A}}$ denotes the eigenvalues of $\B{A}$ in
decreasing order and $\EVV{\B{A}} \in \mathbb{R}^n$ denote the
vector of the ordered eigenvalues, whose $i$th element is
$\EV{i}{}{\B{A}}$. Also, $\EVn{\B{A}}$ denotes the number of
distinct eigenvalues of $\B{A}$, and $\oEV{k}{}{\B{A}}$ and
$\EVm{k}{\B{A}}$, $k=1,2,\ldots,\EVn{\B{A}}$, denote the distinct
eigenvalues of $\B{A}$ in decreasing order and its multiplicity,
respectively, that is, $\oEV{1}{}{\B{A}}
> \oEV{2}{}{\B{A}}
> \ldots > \oEV{\EVn{\B{A}}}{}{\B{A}}$ and $\sum_{k=1}^\EVn{\B{A}} \EVm{k}{\B{A}}
=n$. Finally, we shall use the notation $\B{X}\in \C^{m \times n}
\sim \CGM{m}{n}{\B{M}}{\B{\Sigma}}{\B{\Psi}}$ to denote that a
random matrix $\B{X}$ is (matrix-variate) Gaussian distributed
with the probability density function (pdf)
\begin{align}
    \PDF{\B{X}}{\B{X}}
        =   \pi^{-mn}
            \det\left(\B{\Sigma}\right)^{-n}
            \det\left(\B{\Psi}\right)^{-m}
            \etr\left\{
                    -\B{\Sigma}^{-1}
                    \left(\B{X}-\B{M}\right)
                    \B{\Psi}^{-1}
                    \left(\B{X}-\B{M}\right)^\dag
            \right\}
\EqE
\end{align}
where $\B{M} \in \C^{m \times n}$, $\B{\Sigma}=\B{\Sigma}^\dag \in
\C^{m \times m}>0$, and $\B{\Psi}=\B{\Psi}^\dag \in \C^{n \times
n}>0$.

\section{Signal and Channel Models}  \label{sec:Sec:2}

We consider a MIMO system with $\nTx$ transmit and $\nRx$ receive
antennas, where the channel remains constant for $\Tc$ symbol
periods and changes independently to a new value for each
coherence time, i.e., every $\Tc$ symbols. Since the propagation
coefficients independently acquire new values for every coherence
interval, the channel is memoryless when considering a block
length of $\Tc$ symbols as one channel use with input and output
signals of dimension $\nTx \times \Tc$ and $\nRx
\times \Tc$, respectively. 

\begin{figure}[t]
    \centerline{\includegraphics[width=0.95\textwidth]{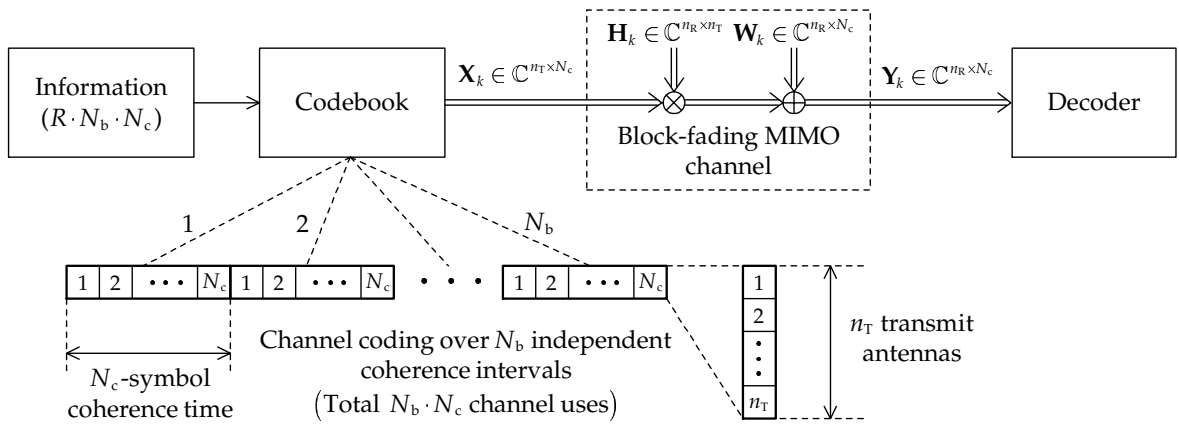}}
    \caption{
        A wireless communication link with $\nTx$ transmit and
        $\nRx$ receive antennas to communicate at a rate $R$ over $\Nb$
        independent $\Tc$-symbol coherence intervals.
    }
    \label{fig:1}
\end{figure}

For an observation interval of $\Nb \Tc$ symbol periods, the
received signal is a sequence $\left\{\B{Y}_k\right\}_{k=1}^\Nb$,
each $\B{Y}_k \in \C^{\nRx \times \Tc}$ is given by
\begin{align}   \label{eq:RS}
    \B{Y}_{k}
    =
        \B{H}_k
        \B{X}_k
        +
        \B{W}_k
        \, ,
    \qquad
    k=1,2,\ldots,\Nb
\end{align}
where $\B{X}_k \in \C^{\nTx \times \Tc}$ are the transmitted
signal matrices, $\B{H}_k \in \C^{\nRx \times \nTx}$ are the
channel matrices, and $\B{W}_k \sim \CGM{\nRx}{\Tc}{\B{0}}{N_0
\B{I}_\nRx}{\B{I}_\Tc}$ are the additive white Gaussian noise
(AWGN) matrices. Fig.~\ref{fig:1} shows a communication link with
$\nTx$ transmit and $\nRx$ receive antennas to communicate at a
rate $R$ (in bits or nats per symbol) over $\Nb$ independent
$\Tc$-symbol coherence intervals. Since the channel is memoryless
with identical channel statistics for each coherence time
interval, the index $k$ can be dropped.

Let $\PDF{\B{X}}{\B{X}}$ be the input probability assignment for
$\B{X} \in \C^{\nTx \times \Tc}$ with covariance $\cov
\bigl\{\vecOp \bigl( \B{X}^\dag \bigr) \bigr\}= \B{Q}^T \otimes
\B{I}_{\Tc}$ subject to the average power constraint of the form
\begin{align}   \label{eq:PC}
    \frac{1}{\Tc} \,
    \EX{\tr\bigl(\B{XX}^\dag\bigr)}
    &=
        \frac{1}{\Tc} \,
        \tr\left(
                    \B{Q}^T
                    \otimes
                    \B{I}_\Tc
           \right)
    =
        \tr\left(\B{Q}\right)
        \leq
        \mathcal{P}
\end{align}
where $\B{Q}$ is the $\nTx \times \nTx$ positive semidefinite
matrix and  $\mathcal{P}$ is the total transmit power over $\nTx$
transmit antennas.  Taking into account spatial fading correlation
at both the transmitter and the receiver, we consider the channel
matrix $\B{H}$ is given by \cite{SFGK:00:COM,CTKV:02:IT}
\begin{align}   \label{eq:H}
    \B{H}
    =
        \RxCM^{1/2}
        \B{H}_0
        \TxCM^{1/2}
\end{align}
where $\TxCM \in \C^{\nTx \times \nTx}>0$ and $\RxCM \in \C^{\nRx
\times \nRx}>0$ are the transmit and receive correlation matrices,
respectively, and $\B{H}_0 \sim
\CGM{\nRx}{\nTx}{\B{0}}{\B{I}_\nRx}{\B{I}_\nTx}$ is a matrix with
i.i.d., zero-mean, unit-variance, complex Gaussian entries. The
$\left(i,j\right)$ entry $H_{ij}$, $i=1,2,\ldots,\nRx$,
$j=1,2,\ldots,\nTx$, of $\B{H}$ is a complex propagation
coefficient between the $j$th transmit antenna and the $i$th
receive antenna with $\EX{\left|H_{ij}\right|^2}=1$. Note that
$\B{H} \sim \CGM{\nRx}{\nTx}{\B{0}}{\RxCM}{\TxCM}$
\cite{SL:03:IT}. With perfect CSI at the receiver, we have the
transition pdf
\begin{align}   \label{eq:TPDF}
\EqF
    p\left(
            \left.\B{Y}\right|
            \B{X},
            \B{H}
     \right)
    =
        \left(
                \pi
                N_0
        \right)^{-\nRx \Tc}
                    \etr\left\{
                                -
                                \frac{1}{N_0}
                                \left(
                                        \B{Y}
                                        -\B{HX}
                                \right)
                                \left(
                                        \B{Y}
                                        -\B{HX}
                                \right)^\dag
                        \right\}
\EqE
\end{align}
which completely characterizes a block-fading MIMO channel.

In what follows, we define the random matrix $\MHH \in \C^{\nXmin
\times \nXmin}>0$ as
\mySepR
\begin{salign}
    \MHH
    \triangleq
        \begin{cases}
            \B{HH}^\dag,     &   \text{if $\nRx \leq \nTx$}
            \\
            \B{H}^\dag \B{H},    &   \text{otherwise}
        \end{cases}
\end{salign}
\mySepR
which is a matrix quadratic form in complex Gaussian matrices,
denoted by $\MHH \sim \CQM{m}{n}{\B{I}_n}{\CMmin}{\CMmax}$
\cite{SL:03:IT}, where
$\nXmin\triangleq\min\left\{\nTx,\nRx\right\}$,
$\nXmax\triangleq\max\left\{\nTx,\nRx\right\}$, and
\begin{salign}
    \left(\CMmin \in \C^{m \times m},\CMmax \in \C^{n \times n}\right)
    =
        \begin{cases}
            \left(\RxCM,\TxCM\right),   &   \text{if $\nRx \leq \nTx$}
            \\
            \left(\TxCM,\RxCM\right),   &   \text{otherwise}.
        \end{cases}
\end{salign}
The pdf of $\MHH \sim \CQM{m}{n}{\B{I}_\nXmax}{\CMmin}{\CMmax}$ is
given by \cite{SWLC:05:WCOM}
\begin{align}   \label{eq:PDF:MHH}
\EqF
    \PDF{\MHH}{\MHH}
    &=
        \frac{1}
             {
                \tilde{\Gamma}_\nXmin\left(\nXmax\right)
             }
        \det\left(
                    \CMmin
            \right)^{-\nXmax}
        \det\left(
                    \CMmax
            \right)^{-\nXmin}
        \det\left(\MHH\right)^{\nXmax-\nXmin}
        \matHyperPFQII{0}{0}{\nXmax}{-\CMmin^{-1}\MHH,\CMmax^{-1}},
        \quad
        \MHH>0,
\EqE
\end{align}
where $\tilde{\Gamma}_m \left(\alpha\right)=\pi^{m\left(m-1
\right)/2}\prod_{i=0}^{m-1} \Gamma\left(\alpha-i\right)$,
$\Re\left\{\alpha\right\}>m-1$, is the complex multivariate gamma
function, $\Gamma\left(\cdot\right)$ is the Euler gamma function,
and $\matHyperPFQII{p}{q}{n}{\cdot}$ is the hypergeometric
function of two Hermitian matrices, defined by
\cite[eq.~(88)]{Jam:64:AMS}.


\section{MIMO Exponent: Reliability--Rate Tradeoff}  \label{sec:Sec:3}

This section is based on Gallager's random coding bound on the
error probability of maximum-likelihood (ML) decoding for a
channel with continuous inputs and outputs \cite{Gal:68:Book}.
Notably, the bound determines the behavior of the error
probability as a function of the rate and the codeword length.
Hence, by determining Gallager's exponent, we can obtain
significant insight into the reliability--rate tradeoff in
communication over MIMO channels and the required codeword length
to achieve a certain level of reliable communication. In
particular, the diversity--multiplexing tradeoff of MIMO channels
\cite{Win:87:JSAC,WSG:94:COM,ZT:03:IT} is a special case of the
reliability--rate tradeoff as the SNR goes to infinity.

\subsection{Random Coding Exponent}

Using the formulation developed in \cite[ch.~7]{Gal:68:Book}, we
obtain the random coding bound on the error probability of ML
decoding over block-fading MIMO channels as\footnote{
        When $\B{X}=\left(X_{ij}\right)$ is an $m \times n$ matrix of complex variables that do not depend functionally on each other,
$
            d\B{X}
            =
                \prod_{i=1}^m
                \prod_{j=1}^n
                d\Re X_{ij} \,
                d\Im X_{ij}.
$
        %
        %
        If $\B{X}\in \C^{m \times m}$ is Hermitian, then
$            d\B{X}
            =
                \prod_{i=1}^m
                    dX_{ii}
                \prod_{i<j}^m
                    d\Re X_{ij} \,
                    d\Im X_{ij}
$.
        }
\begin{align}   \label{eq:RCB}
    P_\text{e}
    \leq
        \left(
                \frac{2e^{r\delta}}{\xi}
        \right)^2
        e^{
                    -\Nb \Tc
                    \,
                    \Er\left(
                                \PDF{\B{X}}{\B{X}},
                                R,
                                \Tc
                       \right)
        }
\end{align}
where $r,\delta \geq 0$ and
\begin{gather}   \label{eq:VA}
    \xi
    \approx
        \frac{\delta}
             {
                \sqrt{2\pi \Nb \sigma_\xi^2}
             }
%
%
    \\
            \label{eq:SV}
    \sigma_\xi^2
    =
        \int_{\B{X}}
            \left[
                \tr\left(\B{XX}^\dag\right)
                -\Tc \mathcal{P}
            \right]^2
            \PDF{\B{X}}{\B{X}}
            d\B{X}.
\EqE
\end{gather}
The random coding exponent
$\Er\left(\PDF{\B{X}}{\B{X}},R,\Tc\right)$ in \eqref{eq:RCB} is
given by
\begin{align}   \label{eq:Er1}
\EqF
    \Er\left(\PDF{\B{X}}{\B{X}},R,\Tc\right)
    =
        \max_{0 \leq \rho \leq 1}
        \left\{
        \max_{r \geq 0}
            E_0\left(\PDF{\B{X}}{\B{X}},\rho,r,\Tc\right)
            -\rho R
        \right\}
\end{align}
with
\begin{align}   \label{eq:Eo}
    &
    E_0\left(\PDF{\B{X}}{\B{X}},\rho,r,\Tc\right)
    \nonumber \\
    & ~~
    =
        -\frac{1}{\Tc}
        \ln\left\{
                    \int_{\B{H}}
                        \PDF{\B{H}}{\B{H}}
                        \int_{\B{Y}}
                            \left(
                                    \int_{\B{X}}
                                        \PDF{\B{X}}{\B{X}}
                                        e^{
                                            r
                                            \left[
                                                    \tr\left(
                                                                \B{XX}^\dag
                                                        \right)
                                                    -\Tc \mathcal{P}
                                            \right]
                                          }
                                        p\left(
                                                \left. \B{Y} \right|
                                                \B{X},
                                                \B{H}
                                          \right)^{1/\left(1+\rho\right)}
                                    d\B{X}
                            \right)^{1+\rho}
                        d\B{Y}
                    d\B{H}
            \right\}.
\end{align}
The parameter $r$ to be optimized may be viewed as a Lagrange
multiplier corresponding to the input-power constraint
\cite{AH:99:TM}.

\subsubsection{Capacity-Achieving Input Distribution}

As in
\cite{Gal:68:Book,Eri:70:IT,Ahm:97:PhD,AM:99:IT,AH:99:TM,KS:95:AEU},
we choose the capacity-achieving distribution for
$\PDF{\B{X}}{\B{X}}$ satisfying the power constraint
\eqref{eq:PC}, namely, 
%
%
\begin{align}   \label{eq:GIPDF}
\EqF
    \PDF{\B{X}}{\B{X}}
    =
        \pi^{-\nTx\Tc}
        \det\left(\B{Q}\right)^{-\Tc}
        \etr\left(
                    -\B{Q}^{-1}
                     \B{XX}^\dag
            \right)
\EqE
\end{align}
with $\tr\left(\B{Q}\right) \leq \mathcal{P}$.\footnote{
        In general, optimization of the input distribution $\PDF{\B{X}}{\B{X}}$ under
        the power constraint \eqref{eq:PC} to maximize the error exponent
        (i.e., to minimize the upper bound) is a difficult task.
        }
Although this choice of the Gaussian input distribution is optimal
only if the rate approaches the channel capacity, it makes the
problem analytically tractable \cite{Gal:68:Book}.

\begin{proposition}       \label{thm:EoGI}

Let $\Eg\left(\B{Q},\rho,r,\Tc\right)$ be
$E_0\left(\PDF{\B{X}}{\B{X}},\rho,r,\Tc\right)$ in \eqref{eq:Eo}
for the Gaussian input distribution $\PDF{\B{X}}{\B{X}}$ of
\eqref{eq:GIPDF}. Then, we have
\begin{align}   \label{eq:EoGI}
\EqF
    \Eg\left(\B{Q},\rho,r,\Tc\right)
    &=
        r
        \mathcal{P}
        \left(1+\rho\right)
        +
        \left(1+\rho\right)
        \ln
        \det\left(\B{I}_\nTx-r\B{Q}\right)
    \nonumber \\
    & \quad
        -\frac{1}{\Tc}
        \ln
        \EX{
            \det\left(
                        \B{I}_\nRx
                        +
                        \frac{
                                \B{H}
                                \left(
                                        \B{Q}^{-1}
                                        -r\B{I}_\nTx
                                \right)^{-1}
                                \B{H}^\dag
                             }
                             {
                                N_0
                                \left(1+\rho\right)
                             }
                 \right)^{-\Tc\rho}
            }.
\EqE
\end{align}

\begin{proof}
See Appendix~\ref{sec:Appendix:I}.
\end{proof}

\end{proposition}

For the case of equal power allocation to each of transmit
antennas, i.e., $\B{Q}=\frac{\mathcal{P}}{\nTx}\B{I}_\nTx$
(because the transmitter has no channel knowledge),
\eqref{eq:EoGI} becomes
\begin{align}       \label{eq:EoGIEP}
    \Eg\left(
                \tfrac{\mathcal{P}}{\nTx}\B{I}_\nTx,
                \rho,
                r,
                \Tc
       \right)
    &=
        r
        \mathcal{P}
        \left(1+\rho\right)
        +
        \nTx
        \left(1+\rho\right)
        \ln\left(
                    \frac{\nTx-r\mathcal{P}}{\nTx}
           \right)
    \nonumber \\
    & \quad
        -\frac{1}{\Tc}
        \ln
        \EX{
            \det\left(
                        \B{I}_\nRx
                        +
                        \frac{
                                \snr
                                \B{HH}^\dag
                             }
                             {
                                \left(\nTx-r\mathcal{P}\right)
                                \left(1+\rho\right)
                             }
                \right)^{-\Tc\rho}
            }
\EqE
\end{align}
where $\snr=\mathcal{P}/N_0$ is the average SNR at each receive
antenna. Let us introduce a new variable $\beta=\nTx-r\mathcal{P}$
where $\beta$ is restricted to the range $0\leq \beta \leq \nTx$
to have a meaningful result in \eqref{eq:EoGIEP}. Then, we have
\begin{align}   \label{eq:EoT}
\EqF
    \EoT{\rho}{\beta}{\Tc}
    &\triangleq
        \left.
            \Eg\left(
                        \tfrac{\mathcal{P}}{\nTx}
                        \B{I}_\nTx,
                        \rho,
                        r,
                        \Tc
               \right)
        \right|_{\beta=\nTx-r\mathcal{P}}
    \nonumber \\
    &=
        \underbrace{
            \left(1+\rho\right)
            \left(\nTx-\beta\right)
            +
            \nTx
            \left(1+\rho\right)
            \ln\left(\beta/\nTx\right)
        }_{
            \triangleq \,
                \EoConst{\rho}{\beta}
           }
        -\frac{1}{\Tc}
         \ln
         \EoRF{\rho}{\beta}{\Tc}{}
\end{align}
where
\begin{align}
            \label{eq:EoRF}
    \EoRF{\rho}{\beta}{\Tc}{}
    =
        \EX{
            \det\left(
                        \B{I}_\nXmin
                        +
                        \frac{
                                \snr
                                \MHH
                             }
                             {
                                \beta
                                \left(1+\rho\right)
                             }
                \right)^{-\Tc\rho}
           }.
\EqE
\end{align}
With maximization over $\beta \in \left[0,\nTx\right]$ and $\rho
\in \left[0,1\right]$ to obtain the tightest bound, we have the
random coding exponent for Gaussian codebooks and equal power
allocation as follows:\footnote{
        The random coding bound can be
        improved by expurgating ``bad" codewords from the code ensemble
        at low rates (see, e.g, \cite{Gal:68:Book}). More details
        for the expurgated exponent of block-fading MIMO channels can be found
        in \cite{Shi:04:PhD}.
        }
\begin{align}   \label{eq:ErGI}
\EqF
    \Er\left(R,\Tc\right)
    &\triangleq
        \Er\left(\PDF{\B{X}}{\B{X}},R,\Tc\right)
            \Bigr|_{
                        \B{X}
                        \sim
                        \CGM{\nTx}{\Tc}{\B{0}}{\frac{\mathcal{P}}{\nTx}\B{I}_\nTx}{\B{I}_\Tc}
                   }
    \nonumber \\
    &=
        \max_{0\leq \rho \leq 1}
        \left\{
        \max_{0\leq \beta \leq \nTx}
            \EoT{\rho}{\beta}{\Tc}
            -\rho R
        \right\}.
\EqE
\end{align}

\mySep

\begin{proposition}     \label{thm:OB}

Let $\OB{\rho}$ be the value of $\beta$ that maximizes
$\EoT{\rho}{\beta}{\Tc}$ defined in \eqref{eq:EoT} for each $\rho
\in \left[0,1\right]$. Then, $\OB{\rho}$ is the solution of
$\partial \EoT{\rho}{\beta}{\Tc}/
\partial \beta =0$ and is always in the range $0<\beta \leq \nTx$.

\begin{proof}
See Appendix~\ref{sec:Appendix:II}.
\end{proof}

\end{proposition}

\mySep

It can be shown using \eqref{eq:DZo} and \eqref{eq:DBEoT} in
Appendix~\ref{sec:Appendix:II} that as $\snr \to \infty$ or $\snr
\to 0$, the optimal value of $\beta$ does not depend on $\Tc$,
that is,
\begin{align*}
\EqF
    \lim_{\snr \to \infty}
    \OB{\rho}
    =
        \nTx
        -
        \frac{
                \nXmin\rho
             }
             {
                1+\rho
             }
    \quad
    \text{and}
    \quad
    \lim_{\snr \to 0}
    \OB{\rho}
    =
        \nTx.
\EqE
\end{align*}
According to Proposition~\ref{thm:OB} and using the general
relation $d\Er\left(R,\Tc\right)/dR=-\rho$, the maximization of
the exponent in \eqref{eq:ErGI} over $\beta \in
\left[0,\nTx\right]$ and $\rho \in \left[0,1\right]$ can be
performed by the following parametric equations:
\begin{align}
    \label{eq:ErGIO}
    \Er\left(R,\Tc\right)
    &=
        \EoT{\rho}{\OB{\rho}}{\Tc}
        -\rho R
    \\
    \label{eq:RO}
    R
    &=
        \left.
        \left[
                \frac{
                        \partial
                        \EoT{\rho}{\beta}{\Tc}
                     }
                     {
                        \partial
                        \rho
                     }
        \right]
        \right|_{\beta=\OB{\rho}}
\end{align}
with
\begin{align}   \label{eq:EoTDR}
    \frac{
            \partial
            \EoT{\rho}{\beta}{\Tc}
         }
         {
            \partial
            \rho
         }
    =
        \underbrace{
            \left(\nTx-\beta\right)
            +
            \nTx
            \ln\left(\beta/\nTx\right)
        }_{
            \triangleq \,
            \dREoConst{\rho}{\beta}
            =
                \frac{
                        \partial
                        \EoConst{\rho}{\beta}
                    }
                    {
                        \partial
                        \rho
                    }
          }
        -
        \frac{1}{\Tc}
        \EoRF{\rho}{\beta}{\Tc}{-1}
        \frac{
                \partial
                \EoRF{\rho}{\beta}{\Tc}{}
             }
             {
                \partial
                \rho
             }
\end{align}
where
\begin{align}       \label{eq:LoTDR}
\EqF
    \frac{
            \partial
            \EoRF{\rho}{\beta}{\Tc}{}
            }
            {
            \partial
            \rho
            }
    &=
        \E\biggl\{
            \Tc
            \det\bigl(
                        \tfrac{1}{\beta}
                        \,
                        \B{\Omega}_{\rho,\beta}
                \bigr)^{-\Tc\rho}
            \left[
                \tfrac{\rho\snr}{\beta\left(1+\rho\right)^2}
                \tr\left\{
                            \MHH
                            \bigl(
                                \tfrac{1}{\beta}
                                \,
                                \B{\Omega}_{\rho,\beta}
                            \bigr)^{-1}
                    \right\}
                -\ln\det\bigl(
                        \tfrac{1}{\beta}
                        \,
                        \B{\Omega}_{\rho,\beta}
                        \bigr)
            \right]
        \biggr\}.
\EqE
\end{align}

\subsubsection{Key Quantities}

The values of $R$ in \eqref{eq:RO} at $\rho =1$ and $\rho=0$ are
the \emph{critical rate} $\Rcr$ and the \emph{ergodic capacity}
$\ErC$ of the channel, respectively
\cite{Gal:68:Book,Eri:70:IT,Ahm:97:PhD,AM:99:IT}. From $\partial
\EoT{\rho}{\beta}{\Tc}/\partial\beta$ in \eqref{eq:DBEoT}, we see
that $\OB{0}=\nTx$ and hence, the ergodic capacity can be written
as
\begin{align}   \label{eq:ECFRCE:1}
\EqF
    \ErC
    &=
        \left.
        \left[
                \frac{
                        \partial
                        \EoT{\rho}{\beta}{\Tc}
                     }
                     {
                        \partial
                        \rho
                     }
        \right]
        \right|_{\rho=0,\,\beta=\nTx}
    \\
            \label{eq:ECFRCE}
    &=
        \EX{
            \ln
            \det\left(
                        \B{I}_\nXmin
                        +
                        \tfrac{\snr}{\nTx}
                        \MHH
                \right)
            }.
\EqE
\end{align}
We remark that the capacity expression \eqref{eq:ECFRCE} obtained
from the exponent is independent of the channel coherence time
$\Tc$ and is in agreement with the previous result
\cite{Fos:96:BLTJ,FG:98:WPC,Tel:99:ETT}. Also, the quantity $E_0$
is defined as the value of the exponent $\Er\left(R,\Tc\right)$ at
$R=0$, referred to as the \textit{exponential error-bound
parameter} \cite{Eri:70:IT,Ahm:97:PhD}, and is given by
$\EoT{1}{\OB{1}}{\Tc}$. This quantity is equal to the value of $R$
at which the exponent becomes zero by setting $\rho=1$ and
$\beta=\OB{1}$. If setting $r=0$ or equivalently $\beta=\nTx$
(i.e., without the constraint on the minimum energy of the
codewords) in \eqref{eq:Eo}, $E_0$ becomes equal to the
\emph{cutoff rate} $R_0$ of the channel
\begin{align}   \label{eq:Ro:1}
\EqF
    R_0
    &=
        \EoT{1}{\nTx}{\Tc}
    \\
        \label{eq:Ro}
    &=
        -\frac{1}{\Tc}
        \ln
        \EX{
            \det\left(
                        \B{I}_\nXmin
                        +
                        \tfrac{\snr}{2\nTx}
                        \MHH
                \right)^{-\Tc}
            }.
\EqE
\end{align}
This is an important parameter, as it determines both the
magnitude of the zero-rate exponent and the rate regime in which
the error probability can be made arbitrarily small by increasing
the codeword length.

\subsubsection{Effect of Channel Coherence---Channel-Incurable Effect}


Using Jensen's inequality, it is easy to show
\begin{align}   \label{eq:JI}
\EqF
    \frac{1}{\Tc}
    \ln
    \EoRF{\rho}{\beta}{\Tc}{}
    \geq
        \frac{1}{\Tc-1}
        \ln \EoRF{\rho}{\beta}{\Tc-1}{}
\end{align}
yielding
\begin{align}   \label{eq:EoTc}
    \EoT{\rho}{\beta}{\Tc}
    \leq
        \EoT{\rho}{\beta}{\Tc-1}.
\EqE
\end{align}
%
%
Therefore, for fixed $R$, the random coding exponent decreases
with $\Tc$, while the channel capacity is independent of $\Tc$.
This reliability reduction is due to the fact that the increase in
$\Tc$ results in a decrease in the number of independent channel
realizations across the code and hence, reduces the effectiveness
of channel coding to mitigate unfavorable fading. We call this
effect of the channel coherence time on communication reliability
``a channel-incurable effect". In particular, since $\lim_{\Tc \to
\infty}
                \frac{1}{\Tc}
                \ln
                \EoRF{\rho}{\beta}{\Tc}{}
            =0$,
we have
%
%
\begin{align}   \label{eq:EoTITc}
\EqF
    \lim_{\Tc\to\infty}
        \EoT{\rho}{\beta}{\Tc}
    =
        \EoConst{\rho}{\beta}
\EqE
\end{align}
leading to $\lim_{\Tc\to\infty} \OB{\rho} = \nTx$ and
$\lim_{\Tc\to\infty} \Er\left(R,\Tc\right) =0$.
%
%
%
%
Therefore, if $\Tc \to \infty$, it is impossible to transmit
information at any positive rate with arbitrary reliability even
with the use of multiple antennas. In fact, $\nTx$ must also
increase without limit so that the so-called \emph{space--time
autocoding effect} takes place, which makes arbitrarily reliable
communications possible \cite{HMH:01:IT}.

\subsection{Evaluation of $\EoT{\rho}{\beta}{\Tc}$, $\partial\EoT{\rho}{\beta}{\Tc}/\partial\beta$, and
$\partial\EoT{\rho}{\beta}{\Tc}/\partial\rho$}

To calculate the random coding exponent, the quantities
$\EoT{\rho}{\beta}{\Tc}$,
$\partial\EoT{\rho}{\beta}{\Tc}/\partial\beta$, and
$\partial\EoT{\rho}{\beta}{\Tc}/\partial\rho$ need to be
determined. We now evaluate them in the following theorem which
will be proven in the next section.

\mySep

\begin{theorem}     \label{thm:RCEE}

Let $\B{H} \sim \CGM{\nRx}{\nTx}{\B{0}}{\RxCM}{\TxCM}$ or $\MHH
\sim \CQM{\nXmin}{\nXmax}{\B{I}_\nXmax}{\CMmin}{\CMmax}$. Then,

\mySep

\begin{enumerate}

\item

$\EoT{\rho}{\beta}{\Tc}$ is given by
\begin{salign}   \label{eq:EoT:f}
    \EoT{\rho}{\beta}{\Tc}
    &=
        \begin{cases}
            \EoConst{\rho}{\beta}
            -
            \frac{1}{\Tc}
            \ln\left(
                    \Kc^{-1} \,
                    \det
                                \begin{bmatrix}
                                    \CM{\nXmin-\Tc\rho}{\CMmin}
                                    \\
                                    \B{\Xi}\left(\rho,\beta\right)
                                \end{bmatrix}
                \right),
            &
            \text{if $\Tc\rho \in \left\{1,2,\ldots,\nXmin\right\}$}
            \\[0.7cm]
            \EoConst{\rho}{\beta}
            +
            \frac{
                    \mathcal{T}_\text{A}
                }{\Tc}
            \ln\left(
                        \frac{\snr}{\beta\left(1+\rho\right)}
                \right)
            &
            \\
                \hspace{1cm}
                -
                \frac{1}{\Tc}
                \ln\left(
                        \mathcal{T}_\text{B}\left(\rho,\Tc\right)
                        \,
                        \det
                                \begin{bmatrix}
                                \CM{\nXmax-\nXmin}{\CMmax}
                                \\
                                \B{\Upsilon}\left(\rho,\beta\right)
                                \end{bmatrix}
                    \right),
            &
            \text{otherwise}.
        \end{cases}
\end{salign}
If $\TxCM=\B{I}_\nTx$ and $\RxCM=\B{I}_\nRx$ (i.i.d.\ MIMO
channel), then $\EoT{\rho}{\beta}{\Tc}$ reduces to
\begin{align}   \label{eq:EoT:IID:f}
\EqF
    \EoT{\rho}{\beta}{\Tc}
    =
        \EoConst{\rho}{\beta}
        -
        \frac{1}{\Tc}
        \ln\Bigl(
                \Kiid^{-1} \,
                \det
                    \B{\Upsilon}_\text{iid}\left(\rho,\beta\right)
            \Bigr).
\EqE
\end{align}

\item

$\partial\EoT{\rho}{\beta}{\Tc}/\partial\beta$ is given by
\mySepR
\begin{salign}
    \frac{
            \partial
            \EoT{\rho}{\beta}{\Tc}
        }{
            \partial
            \beta
        }
    =
        \dBEoConst{\rho}{\beta}
        -
        \frac{
                \mathcal{T}_\text{A}
            }{
                \Tc
                \beta
            }
        -
        \frac{1}{\Tc}
        \tr\left\{
                    \begin{bmatrix}
                        \CM{\nXmax-\nXmin}{\CMmax}
                        \\
                        \B{\Upsilon}\left(\rho,\beta\right)
                    \end{bmatrix}^{-1}
                    \begin{bmatrix}
                        \B{0}
                        \\
                        \B{\Upsilon}^{\left(\beta\right)}\left(\rho,\beta\right)
                    \end{bmatrix}
            \right\}.
\end{salign}
\mySepR
If $\TxCM=\B{I}_\nTx$ and $\RxCM=\B{I}_\nRx$, then
\begin{align}
\EqF
    \frac{
            \partial
            \EoT{\rho}{\beta}{\Tc}
        }{
            \partial
            \beta
        }
    =
        \dBEoConst{\rho}{\beta}
        -
        \frac{1}{\Tc}
        \tr\left\{
                    \B{\Upsilon}^{-1}_\text{iid}\left(\rho,\beta\right)
                    \B{\Upsilon}_\text{iid}^{\left(\beta\right)}\left(\rho,\beta\right)
            \right\}.
\EqE
\end{align}

\item

$\partial\EoT{\rho}{\beta}{\Tc}/\partial\rho$ is given by
%
%
%
%
%
%
%
%
%
\mySepR
\begin{salign}
    \frac{
            \partial
            \EoT{\rho}{\beta}{\Tc}
        }{
            \partial
            \rho
        }
    &=
        \dREoConst{\rho}{\beta}
        -
        \frac{
                \mathcal{T}_\text{A}
            }{
                \Tc
                \left(1+\rho\right)
            }
        -
        \sum\limits_{i=1}^{\EVn{\CMmin}}
        \sum\limits_{j=1}^{\EVm{i}{\CMmin}-1}
            \frac{
                    j
                }{
                    \Tc\rho-\nXmin+\EVm{i}{\CMmin}-j
                }
    \nonumber \\
    & \hspace{0.5cm}
        +
        \sum\limits_{k=1}^{\nXmin-1}
            \frac{
                    k
                }{
                    \Tc\rho-k
                }
        -
        \frac{1}{\Tc}
        \tr\left\{
                    \begin{bmatrix}
                        \CM{\nXmax-\nXmin}{\CMmax}
                        \\
                        \B{\Upsilon}\left(\rho,\beta\right)
                    \end{bmatrix}^{-1}
                    \begin{bmatrix}
                        \B{0}
                        \\
                        \B{\Upsilon}^{\left(\rho\right)}\left(\rho,\beta\right)
                    \end{bmatrix}
            \right\},
    \nonumber \\
    & \hspace{7.5cm}
        \text{$\Tc\rho \ne 1,2,\ldots,\nXmin-1$}.
\end{salign}
\mySepR
If $\TxCM=\B{I}_\nTx$ and $\RxCM=\B{I}_\nRx$, then
\begin{align}
\EqF
    \frac{
            \partial
            \EoT{\rho}{\beta}{\Tc}
        }{
            \partial
            \rho
        }
    =
        \dREoConst{\rho}{\beta}
        -
        \frac{1}{\Tc}
        \tr\left\{
                    \B{\Upsilon}^{-1}_\text{iid}\left(\rho,\beta\right)
                    \B{\Upsilon}_\text{iid}^{\left(\rho\right)}\left(\rho,\beta\right)
            \right\}.
\EqE
\end{align}


\end{enumerate}
The quantities $\Kc$, $\Kiid$, $\mathcal{T}_\text{A}$,
$\mathcal{T}_\text{B}\left(\rho,\Tc\right)$, and the matrices
$\CM{\cdot}{\cdot}$, $\B{\Xi}\left(\rho,\beta\right)$,
$\B{\Upsilon}\left(\rho,\beta\right)$,
$\B{\Upsilon}^{\left(\beta\right)}\left(\rho,\beta\right)$,
$\B{\Upsilon}^{\left(\rho\right)}\left(\rho,\beta\right)$,
$\B{\Upsilon}_\text{iid}\left(\rho,\beta\right)$,
$\B{\Upsilon}_\text{iid}^{\left(\beta\right)}\left(\rho,\beta\right)$,
and
$\B{\Upsilon}_\text{iid}^{\left(\rho\right)}\left(\rho,\beta\right)$
are given in Table~\ref{table:M}.

\mySep

\end{theorem}

\mySep

\begin{corollary}[Ergodic Capacity]   \label{cor:ErC}

If $\B{H} \sim \CGM{\nRx}{\nTx}{\B{0}}{\RxCM}{\TxCM}$, then the
ergodic capacity $\ErC$ is given by
\mySepR
\begin{sgather}     \label{eq:ErC}
    \ErC
    =
        \tr\left\{
                    \begin{bmatrix}
                        \CM{\nXmax-\nXmin}{\CMmax}
                        \\
                        \B{\Upsilon}\left(0,\nTx\right)
                    \end{bmatrix}^{-1}
                    \begin{bmatrix}
                        \B{0}
                        \\
                        \B{\Lambda}
                    \end{bmatrix}
            \right\}
        -\left(\nXmin-1\right)
        +
        \sum_{i=1}^{\EVn{\CMmin}}
        \sum_{j=1}^{\EVm{i}{\CMmin}-1}
            \frac{
                    j
                }{
                    \nXmin-\EVm{i}{\CMmin}+j
                }
\intertext{with $\B{\Lambda}
            \in \mathbb{R}^{\nXmin \times \nXmax}$ given by
}
    \B{\Lambda}
    =
        \begin{bmatrix}
            \B{\Lambda}_{1,1}
            &
            \cdots
            &
            \B{\Lambda}_{1,\EVn{\CMmax}}
            \\
            \vdots
            &
            \ddots
            &
            \vdots
            \\
            \B{\Lambda}_{\EVn{\CMmin},1}
            &
            \cdots
            &
            \B{\Lambda}_{\EVn{\CMmin},\EVn{\CMmax}}
        \end{bmatrix}
\end{sgather}
\mySepR
where the $\left(i,j\right)$th entry $\Lambda_{p,q,ij}$ of
$\B{\Lambda}_{p,q} \in \mathbb{R}^{\EVm{p}{\CMmin} \times
\EVm{q}{\CMmax}}$, $p=1,\ldots,\EVn{\CMmin}$,
$q=1,\ldots,\EVn{\CMmax}$, is
\begin{align}
    \Lambda_{p,q,ij}
    =
        \mathcal{G}_{i+j-1,2}
            \left(
                    \tfrac{\snr}{\nTx} \,
                    \oEV{p}{}{\CMmin},
                    \oEV{q}{}{\CMmax},
                    \nXmin-i+1
            \right).
\EqE
\end{align}

\begin{proof}
Note that
\mySepR
\begin{salign}      \label{eq:dREoT:C}
    \tr\left\{
                \begin{bmatrix}
                    \CM{\nXmax-\nXmin}{\CMmax}
                    \\
                    \B{\Upsilon}\left(0,\nTx\right)
                \end{bmatrix}^{-1}
                \begin{bmatrix}
                    \B{0}
                    \\
                    \B{\Upsilon}^{\left(\rho\right)}\left(0,\nTx\right)
                \end{bmatrix}
        \right\}
    &=
        -
        \Tc
        \tr\left\{
                    \begin{bmatrix}
                        \CM{\nXmax-\nXmin}{\CMmax}
                        \\
                        \B{\Upsilon}\left(0,\nTx\right)
                    \end{bmatrix}^{-1}
                    \begin{bmatrix}
                        \B{0}
                        \\
                        \B{\Lambda}
                    \end{bmatrix}
            \right\}
    \nonumber \\
    & \quad
        +
        \underbrace{
            \nTx
                \tr\left\{
                        \begin{bmatrix}
                            \CM{\nXmax-\nXmin}{\CMmax}
                            \\
                            \B{\Upsilon}\left(0,\nTx\right)
                        \end{bmatrix}^{-1}
                        \begin{bmatrix}
                            \B{0}
                            \\
                            \B{\Upsilon}^{\left(\beta\right)}\left(0,\nTx\right)
                        \end{bmatrix}
                    \right\}
        }_{
            = \,
            -\mathcal{T}_\text{A}
            }.
\end{salign}
\mySepR
The proof follows immediately from \eqref{eq:ECFRCE:1},
Theorem~\ref{thm:RCEE}.3 with $\rho=0$ and $\beta=\nTx$, and
\eqref{eq:dREoT:C}.
\end{proof}

\end{corollary}

\mySep

Note that the expression \eqref{eq:ErC} for the ergodic capacity
$\ErC$ is sufficiently general and applicable to arbitrary
structures of correlation matrices $\TxCM$ and $\RxCM$, and hence,
embraces all the previously known results for i.i.d.\ channels
($\TxCM=\B{I}_\nTx$, $\RxCM=\B{I}_\nRx$)
\cite{Tel:99:ETT,SL:03:IT}, one-sided correlated channels
($\CMmin=\B{I}_\nXmin$ \cite{CWZ:03:IT} or $\CMmax=\B{I}_\nXmax$
\cite{SRS:03:IT}), and doubly correlated channels
\cite{SWLC:05:WCOM} (where all the eigenvalues of $\TxCM$ and
$\RxCM$ are assumed to be distinct) as special cases of
\eqref{eq:ErC}.

\mySep \mySep

\begin{corollary}[Cutoff Rate]   \label{cor:cutoff rate}

If $\B{H} \sim \CGM{\nRx}{\nTx}{\B{0}}{\RxCM}{\TxCM}$, then the
cutoff rate $R_0$ is given by
\begin{salign}
    R_0
    =
        \begin{cases}
            -
            \frac{1}{\Tc}
            \ln\left(
                    \Kc^{-1} \,
                    \det
                                \begin{bmatrix}
                                    \CM{\nXmin-\Tc}{\CMmin}
                                    \\
                                    \B{\Xi}\left(1,\nTx\right)
                                \end{bmatrix}
                \right),
            &
            \text{if $\Tc \in
            \left\{1,2,\ldots,\nXmin\right\}$}
            \\[0.7cm]
            \frac{
                    \mathcal{T}_\text{A}
                }{\Tc}
            \ln\left(
                        \frac{\snr}{2\nTx}
                \right)
            -
            \frac{1}{\Tc}
            \ln\left(
                        \mathcal{T}_\text{B}\left(1,\Tc\right)
                        \det
                                    \begin{bmatrix}
                                        \CM{\nXmax-\nXmin}{\CMmax}
                                        \\
                                        \B{\Upsilon}\left(1,\nTx\right)
                                    \end{bmatrix}
                \right),
            &
            \text{otherwise}.
        \end{cases}
\end{salign}
In particular, if $\TxCM=\B{I}_\nTx$ and $\RxCM=\B{I}_\nRx$, then
we have
\begin{align}
\EqF
    R_0
    =
        -
        \frac{1}{\Tc}
        \ln\Bigl(
                    \Kiid^{-1} \,
                \det
                    \B{\Upsilon}_\text{iid}\left(1,\nTx\right)
            \Bigr).
\EqE
\end{align}

\begin{proof}
It follows immediately from \eqref{eq:Ro:1} and
Theorem~\ref{thm:RCEE}.1 with $\rho=1$ and $\beta=\nTx$.
\end{proof}

\end{corollary}

\mySep


\subsection{Coding Requirement}     \label{sec:CR}

As in \cite{AM:99:IT}, we can approximate the required codeword
length to achieve a prescribed error probability $P_\text{e}$ at a
rate $R$ by solving for $\Nb$ in the following equation:
\begin{align}   \label{eq:PeCL}
    P_\text{e}
    =
        \left( 2e^{r\delta}/\xi\right)^2
        e^{
                    -\Nb\Tc
                     \Er\left(R,\Tc\right)
         }.
\EqE
\end{align}
Using \eqref{eq:VA}, it is easy to see that the factor $\left(
2e^{r\delta}/\xi\right)^2$ in \eqref{eq:PeCL} is minimized over
$\delta \geq 0$, for large $\Nb$, by choosing $\delta=1/r$
\cite{Gal:68:Book}. This yields
\begin{align}   \label{eq:LN}
    \min_{\delta \geq 0}
    \left(2e^{r\delta}/\xi\right)^2
    \approx
        8\pi
        e^2
        \sigma_\xi^2
        r^2
        \Nb
    \quad
    \text{for large $\Nb$}.
\EqE
\end{align}
Also, from \eqref{eq:SV} and \cite[Lemma~5]{SW:05:IT}, we have
\begin{align}   \label{eq:SVGI}
    \sigma_\xi^2
    =
        \Tc \mathcal{P}^2/\nTx.
\EqE
\end{align}
Combining \eqref{eq:LN} and \eqref{eq:SVGI} together with the fact
that $\beta =\nTx-r\mathcal{P}$, \eqref{eq:PeCL} can be written as
%
%
%
\begin{align}   \label{eq:PeCLO}
\EqF
    P_\text{e}
    =
        \left(8\pi/\nTx\right)
                \left\{
                        \nTx-
                        \OB{\rho}
                \right\}^2
        \Nb
        \Tc
        \,
        e^{
                    -\Nb\Tc
                     \Er\left(R,\Tc\right)+2
        }.
\EqE
\end{align}
After solving for $\Nb$ in \eqref{eq:PeCLO}, we take $L=\Tc \cdot
\left\lceil \Nb \right\rceil$ as our estimate of the codeword
length (in symbol) required to achieve $P_\text{e}$ at the rate
$R$, where
        $\left\lceil \cdot \right\rceil$
        denotes the smallest integer larger than or equal an enclosed
        quantity.

\section{Proof of the Main Theorem}     \label{sec:proof}

In this section, we provide proofs of the main results stated in
Theorem~\ref{thm:RCEE}. The methodology recently developed in
\cite{SWLC:05:WCOM} and \cite{SW:05:IT} for dealing with random
matrices paves a way to prove the theorem.

\subsection{Proof of Theorem~\ref{thm:RCEE}.1}

%
%
Using the same steps leading to \cite[Theorem~1]{SWLC:05:WCOM}, we
get
\begin{align}       \label{eq:EoRF:1}
\EqF
    \EoRF{\rho}{\beta}{\Tc}{}
    &=
        \int_{\MHH=\MHH^\dag >0}
            \det\left(
                        \B{I}_\nXmin
                        +
                        \rbSnr
                        \MHH
                \right)^{-\Tc\rho}
            \PDF{\MHH}{\MHH}
            d\MHH
    \nonumber \\
    &=
        \frac{
                \pi^{
                        \nXmin
                        \left(\nXmin-1\right)
                    }
                \det\left(
                            \CMmax
                    \right)^{-\nXmin}
             }
             {
                \tilde{\Gamma}_\nXmin\left(\nXmax\right)
                \tilde{\Gamma}_\nXmin\left(\nXmin\right)
             }
        \int_{\EVV{\MHH}}
            \,
            \prod_{k=1}^\nXmin
                \EV{k}{\nXmax-\nXmin}{\MHH}
            \prod_{i<j}^\nXmin
                \bigl(
                \EV{i}{}{\MHH}
                -\EV{j}{}{\MHH}
                \bigr)^2
    \nonumber \\
    & \hspace{2.5cm}
            \times
            \matHyperPFQII{1}{0}{\nXmin}
                          {
                            \Tc\rho;
                            \B{D},
                            -
                            \rbSnr \,
                            \CMmin
                           }
            \matHyperPFQII{0}{0}{\nXmax}
                          {
                            \B{D},
                            -\CMmax^{-1}
                           }
            \,
            d\EVV{\MHH}
\EqE
\end{align}
where $\rbSnr=\frac{\snr}{\beta\left(1+\rho\right)}$ and
$\B{D}=\mathrm{diag}\left(\EV{1}{}{\MHH},\EV{2}{}{\MHH},\ldots,\EV{\nXmin}{}{\MHH}\right)$.\footnote{
    For $\B{A}=\B{A}^\dag \in \C^{p \times p}>0$, we denote
    $\int_{\EVV{\B{A}}} d\EVV{\B{A}} = \int_{0}^{\EV{p-1}{}{\B{A}}}\int_{\EV{p}{}{\B{A}}}^{\EV{p-2}{}{\B{A}}}\cdots\int_{\EV{2}{}{\B{A}}}^{\infty} d\EV{1}{}{\B{A}}d\EV{2}{}{\B{A}} \cdots
    d\EV{p}{}{\B{A}}$. If the integrand is symmetric in
    $\EV{1}{}{\B{A}},\EV{2}{}{\B{A}},\ldots,\EV{p}{}{\B{A}}$, then
    \begin{align*}
        \int_{\EVV{\B{A}}}
        \, d\EVV{\B{A}}
        =
            \frac{1}{p!}
        \underbrace{
            \int_{0}^{\infty}
            \cdots
            \int_{0}^{\infty}
        }_{\text{$p$-fold}}
            \,
            d\EV{1}{}{\B{A}}
            d\EV{2}{}{\B{A}}
            \cdots
            d\EV{p}{}{\B{A}}.
    \end{align*}
%
    }
Successively applying the generic determinantal formula for
hypergeometric functions of matrix arguments
\cite[Lemma~4]{SW:05:IT} and the generalized Cauchy--Binet formula
\cite[Lemma~2]{SWLC:05:WCOM}, the integral in \eqref{eq:EoRF:1}
can be evaluated, after some algebra, as
\mySepR
\begin{salign}      \label{eq:EoRF:f}
    \EoRF{\rho}{\beta}{\Tc}{}
    &=
        \rbSnr^{-\mathcal{T}_\text{A}} \,
        \mathcal{T}_\text{B}\left(\rho,\Tc\right) \,
        \det\left(
                    \begin{bmatrix}
                        \CM{\nXmax-\nXmin}{\CMmax}
                        \\
                        \B{\Upsilon}\left(\rho,\beta\right)
                    \end{bmatrix}
            \right),
    \quad
    \Tc\rho
    \ne 1,2,\ldots,\nXmin-1.
\end{salign}

Substituting \eqref{eq:EoRF:f} into \eqref{eq:EoT} gives the
second case of \eqref{eq:EoT:f}. It should be noted that the
formula in the second case of \eqref{eq:EoT:f} has singular points
at $\Tc\rho =1,2,\ldots,\nXmin-1$ for each $\rho \in
\left(0,1\right]$. These singularities stem from the quantity
$\mathcal{T}_\text{B}\left(\rho,\Tc\right)$, which can be
alleviated using the following analysis.

Suppose that $\Tc \rho$ is a positive integer. Then, using
\cite[Lemma~1]{SW:05:IT}, we have
\mySepR
\begin{align}   \label{eq:EoRF:i}
\EqF
    \EoRF{\rho}{\beta}{\Tc}{}
    &=
        \E_{\B{\MHH}}\Bigl\{
            \E_{\B{S}}\Bigl\{
                \etr\left(
                        -
                        \rbSnr
                        \MHH
                        \B{S}
                        \B{S}^\dag
                    \right)
                \Bigr\}
           \Bigr\}
    \nonumber \\
    &=
        \EXs{\B{S}}{
            \det\left(
                        \B{I}_{\nXmin \nXmax}
                        +
                        \rbSnr
                        \B{S}
                        \B{S}^\dag
                        \CMmin
                        \otimes
                        \CMmax
                \right)^{-1}
           }
    \nonumber \\
    &=
        \EXs{\B{S}}{
            \det\left(
                        \B{I}_{\nXmin \nXmax}
                        +
                        \rbSnr
                        \B{S}^\dag
                        \CMmin
                        \B{S}
                        \otimes
                        \CMmax
                \right)^{-1}
           }
\EqE
\end{align}
\mySepR
where $\B{S} \sim
\CGM{\nXmin}{\Tc\rho}{\B{0}}{\B{I}_\nXmin}{\B{I}_{\Tc\rho}}$ is a
complex Gaussian matrix statistically independent of $\MHH$, and
the last equality follows from the fact that
$\B{S}\B{S}^\dag\CMmin$ and $\B{S}^\dag\CMmin\B{S}$ have the same
nonzero eigenvalues.

If $\Tc \rho \in \left\{1,2,\ldots,\nXmin\right\}$, then
$\B{Z}=\B{S}^\dag \CMmin \B{S} \sim
\CQM{\Tc\rho}{\nXmin}{\B{I}_\nXmin}{\B{I}_{\Tc\rho}}{\CMmin}$.
Hence, using \cite[Theorem~9]{SW:05:IT}, \eqref{eq:EoRF:i} for the
case of $\Tc \rho \in \left\{1,2,\ldots,\nXmin\right\}$ can be
written as
\begin{salign}       \label{eq:EoRF:int}
    \EoRF{\rho}{\beta}{\Tc}{}
    &=
        \EXs{\EVV{\B{Z}}}{
            \prod_{k=1}^{\Tc\rho}
                \det\Bigl\{
                            \B{I}_{\nXmax}
                            +
                            \rbSnr
                            \EV{k}{}{\B{Z}}
                            \,
                            \CMmax
                    \Bigr\}^{-1}
           }
    \nonumber \\
    &=
        \Kc^{-1}
        \int_{\EVV{\B{Z}}}
            \prod_{k=1}^{\Tc\rho}
                \det\Bigl\{
                            \B{I}_{\nXmax}
                            +
                            \rbSnr
                            \EV{k}{}{\B{Z}}
                            \,
                            \CMmax
                    \Bigr\}^{-1}
    \nonumber \\
    & \hspace{2cm}
    \times
            \det\left(
                        \begin{bmatrix}
                            \CM{\nXmin-\Tc\rho}{\CMmin}
                            \\
                            \grave{\B{\Xi}}
                        \end{bmatrix}
                    \right)
            \det_{1\leq i,j \leq \Tc\rho}
                \bigl(
                        \EV{j}{i-1}{\B{Z}}
                \bigr)
            \,
            d\EVV{\B{Z}}
\end{salign}
where
$    \grave{\B{\Xi}}
    =
        \begin{bmatrix}
            \grave{\B{\Xi}}_1
            &
            \grave{\B{\Xi}}_2
            &
            \cdots
            &
            \grave{\B{\Xi}}_{\EVn{\B{Z}}}
        \end{bmatrix}
$
and the $\left(i,j\right)$th entry $\grave{\Xi}_{k,ij}$ of
$\grave{\B{\Xi}}_k \in \mathbb{R}^{\Tc\rho \times
\EVm{k}{\CMmin}}$, $k=1,2,\ldots,\EVn{\CMmin}$, is given by
\begin{align}
\EqF
    \grave{\Xi}_{k,ij}
    =
        \EV{i}{j-1}{\B{Z}}
        e^{-\EV{i}{}{\B{Z}}/\oEV{k}{}{\CMmin}}.
\EqE
\end{align}

Now, applying \cite[Lemma~2]{SWLC:05:WCOM} to \eqref{eq:EoRF:int}
yields
\mySepR
\begin{salign}       \label{eq:EoRF:int:1}
    \EoRF{\rho}{\beta}{\Tc}{}
    &=
        \Kc^{-1} \,
        \det\left(
                    \begin{bmatrix}
                        \CM{\nXmin-\Tc\rho}{\CMmin}
                        \\
                        \B{\Xi}\left(\rho,\beta\right)
                    \end{bmatrix}
                \right)
\end{salign}
\mySepR
where the $\left(i,j\right)$th entry
$\Xi_{k,ij}\left(\rho,\beta\right)$ of the $k$th constituent
matrix $\B{\Xi}_k\left(\rho,\beta\right)$ is given by
\begin{align}   \label{eq:EoRF:int:3}
\EqF
    \Xi_{k,ij}\left(\rho,\beta\right)
    &=
        \int_{0}^{\infty}
            \det\left(
                    \B{I}_{\nXmax}
                    +
                    \rbSnr
                    z
                    \CMmax
                \right)^{-1}
            z^{i+j-2}
            e^{-z/\oEV{k}{}{\CMmin}}
            \,
            dz.
\EqE
\end{align}
Using the \emph{characteristic coefficients}
\cite[Definition~6]{SW:05:IT}, \eqref{eq:EoRF:int:3} can be
written as
\begin{align}   \label{eq:EoRF:int:4}
\EqF
    \Xi_{k,ij}\left(\rho,\beta\right)
    &=
        \sum_{p=1}^{\EVn{\CMmax}}
        \sum_{q=1}^{\EVm{p}{\CMmax}}
            \PFC{p}{q}{\CMmax}
            \int_{0}^{\infty}
                \bigl(
                        1+\rbSnr
                        \oEV{p}{}{\CMmax}
                        z
                \bigr)^{-q}
                z^{i+j-2}
                e^{-z/\oEV{k}{}{\CMmin}}
                \,
                dz
\EqE
\end{align}
where $\PFC{p}{q}{\CMmax}$ is the $\left(p,q\right)$th
characteristic coefficient of $\CMmax$. Finally, substituting
\eqref{eq:EoRF:int:1} into \eqref{eq:EoT} gives the first case of
\eqref{eq:EoT:f} and hence, we complete the proof of the first
part.

\subsection{Proofs of Theorem~\ref{thm:RCEE}.2 and \ref{thm:RCEE}.3}

The second and third parts can be obtained by differentiating
$\EoT{\rho}{\beta}{\Tc}$ in Theorem~\ref{thm:RCEE}.1 with respect
to $\beta$ and $\rho$, respectively, with the help of the
logarithmic derivative of a determinant
\cite[Theorem~9.4]{Lax:97:Book} (or more generally
\cite[Lemma~1]{SWLC:05:WCOM}).

\section{Numerical Results and Discussion}      \label{sec:Sec:4}

In this section, we provide some numerical results to illustrate
the reliability--rate tradeoff in block-fading MIMO channels. For
spatial fading correlation, we consider an exponential correlation
model with $\TxCM=\bigl(\TxCC^{|i-j|}\bigr)$ and
$\RxCM=\bigl(\RxCC^{|i-j|}\bigr)$, $\TxCC,\,\RxCC \in [0,1)$, in
all examples.

\begin{figure}[t]
    \centerline{\includegraphics[width=0.7\textwidth]{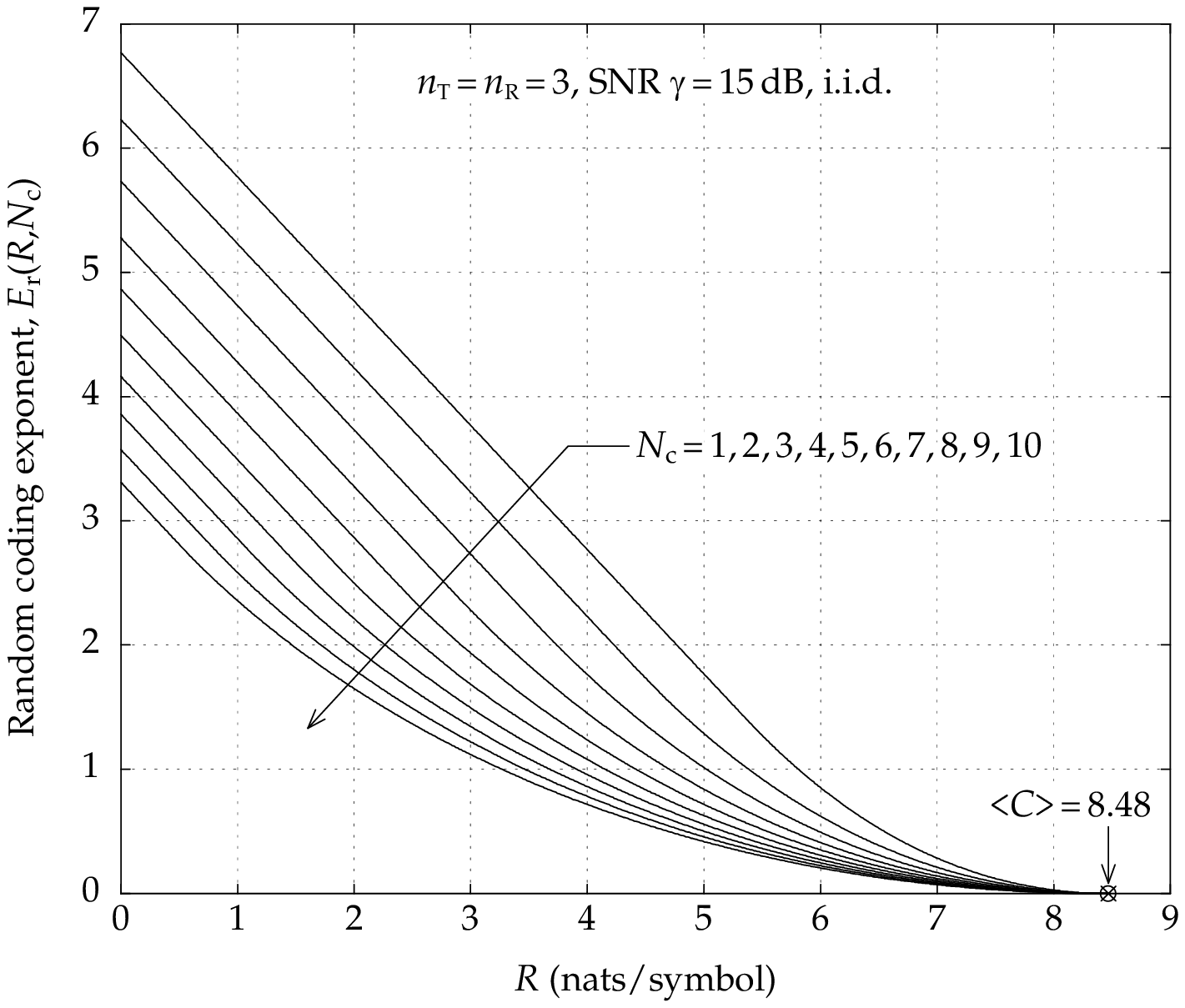}}
    \caption{
        Random coding exponent for i.i.d.\ MIMO channels ($\TxCC=0$, $\RxCC=0$) when $\Tc=1$,
        $2$, $3$, $4$, $5$, $6$, $7$, $8$, $9$, and $10$. $\nTx=\nRx=3$
        and $\snr=15$ dB.
    }
    \label{fig:2}
\end{figure}

\begin{figure}[t]
    \centerline{\includegraphics[width=0.7\textwidth]{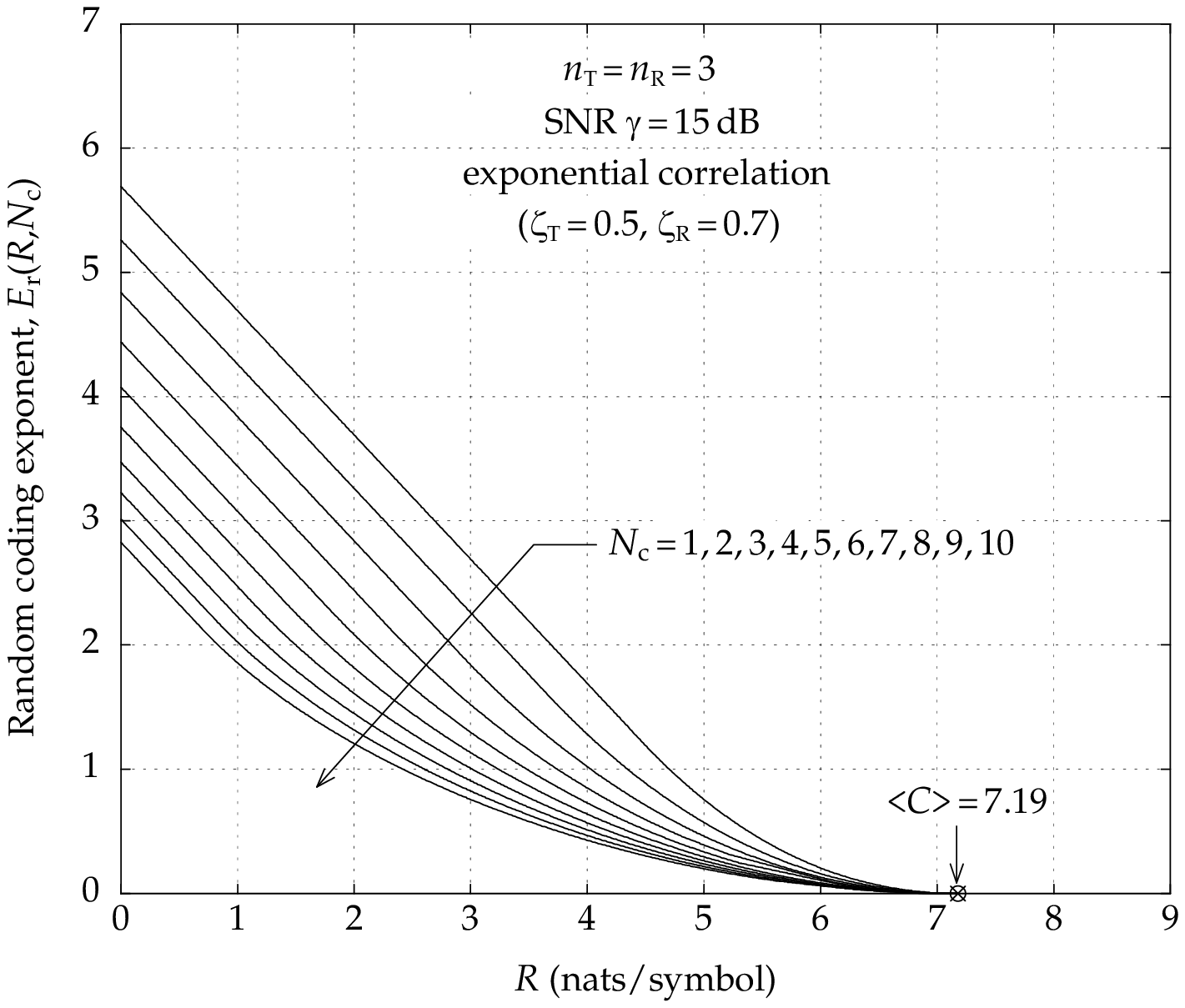}}
    \caption{
        Random coding exponent for exponentially correlated MIMO
        channels with $\TxCC=0.5$ and $\RxCC=0.7$ when $\Tc=1$,
        $2$, $3$, $4$, $5$, $6$, $7$, $8$, $9$, and $10$. $\nTx=\nRx=3$
        and $\snr=15$ dB.
    }
    \label{fig:3}
\end{figure}

To ascertain the effect of the channel coherence on the error
exponent, Figs.~\ref{fig:2} and \ref{fig:3}, respectively, show
the random coding exponent $\Er\left(R,\Tc\right)$ as a function
of a rate $R$ for i.i.d.\ ($\TxCC=0$, $\RxCC=0$) and exponentially
correlated ($\TxCC=0.5$, $\RxCC=0.7$) MIMO channels at $\snr=15$
dB, where $\nTx=\nRx=3$ and $\Tc$ ranges from $1$ to $10$. We can
see from the figures that the exponent at a rate $R$ below the
ergodic capacity decreases with $\Tc$, while the ergodic capacity
remains constant for all $\Tc$ (i.e., $8.48$ and $7.19$
nats/symbol for Figs.~\ref{fig:2} and \ref{fig:3}, respectively).
For example, the error exponents at rates $R \leq \Rcr$ for
$\Tc=10$ are reduced by roughly $3.46$ and $2.86$ for i.i.d.\ and
exponentially correlated cases, respectively, compared with those
for $\Tc=1$. This reduction in the exponent, consequently,
requires using a longer code to achieve the same error
probability. Hence, we see that unlike the capacity (with perfect
receive CSI), the channel coherence time plays a fundamental role
in the error exponent or the reliability of communications.

\begin{figure}[t]
    \centerline{\includegraphics[width=0.7\textwidth]{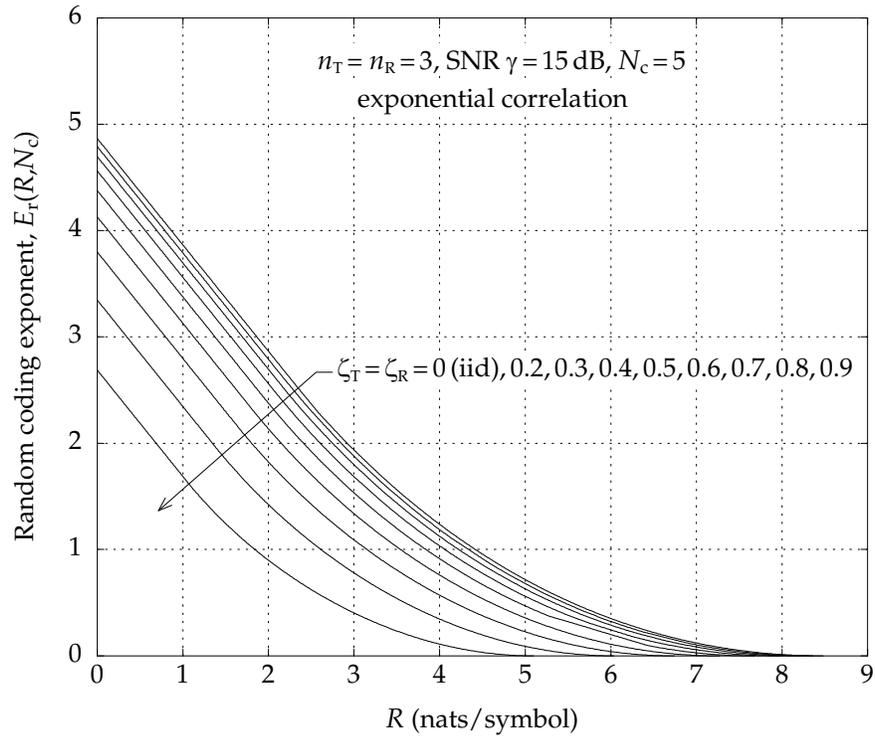}}
    \caption{
        Random coding exponent for exponentially correlated MIMO
        channels when $\TxCC=\RxCC=0$ (i.i.d.), $0.2$, $0.3$, $0.4$, $0.5$, $0.6$,
        $0.7$, $0.8$, and $0.9$. $\nTx=\nRx=3$ and $\snr=15$ dB.
    }
    \label{fig:4}
\end{figure}

Fig.~\ref{fig:4} demonstrates the effect of spatial fading
correlation on the random coding exponent, where $\TxCC=\RxCC
=\zeta$, $\snr=15$ dB, $\nTx=\nRx=3$, $\Tc=5$, and $\zeta$ ranges
from $0$ (i.i.d.) to $0.9$. As seen from the figure, there exists
a remarkable reduction in the exponent at the same rate due to
correlation, especially for $\zeta \geq 0.5$. The amount of
reduction in the exponent at rates $R \leq \Rcr$, relative to the
i.i.d.\ MIMO exponent, ranges from $0.07$ for $\zeta =0.2$ to
$2.17$ for $\zeta=0.9$, indicating that a longer code is required
to achieve the same level of reliable communications.
Equivalently, a decrease in the information rate is required for
more correlated channels to achieve the same value of the
exponent. For example, the exponent at a rate $3$ nats/symbol are
$1.94$ and $1.53$ for the i.i.d.\ and correlated
($\TxCC=\RxCC=0.5$) channels, respectively. This implies that
$27\%$ increase in the codeword length, due to spatial fading
correlation, is required to achieve a rate $3$ nats/symbol with
the same communication reliability.

To get more insight into the influences of the number of antennas,
channel coherence time, and fading correlation on a coding
requirement for MIMO channels, the codeword length required to
achieve $P_\text{e} \leq 10^{-6}$ at a rate $8.0$ bits/symbol
($5.55$ nats/symbol) are investigated in Tables~
\ref{table:1}--\ref{table:3}. The codeword lengths in the tables
are calculated in such a manner as described in
Section~\ref{sec:CR}. Table~\ref{table:1} serves to demonstrate
the effect of increasing the number of antennas on the coding
requirement, in which the required codeword length $L$ is shown
for i.i.d.\ MIMO channels with $\Tc=5$. Note that it is impossible
to reliably communicate at a rate $8.0$ bits/symbol below the SNR
$\snr$ of $14.55$ dB, $9.68$ dB, and $6.79$ dB for $\nTx=\nRx=2$,
$3$, and $4$, respectively, since these SNR's are required to
attain the ergodic capacity $\ErC$ of $8.0$ bits/symbol in each of
the cases. As seen from the table, with increasing the number of
antennas at both transmit and receive sides, the required codeword
lengths are remarkably reduced. This is due to the advantages of
the use of multiple antennas, e.g., spatial multiplexing and
diversity gains \cite{SL:03:IT}. For example, at $\snr=16$ dB,
increasing the number of antennas at both sides from $2$ to $3$
and $4$ reduces the corresponding codeword length to almost
$2.8{\%}$ and $0.9{\%}$ of the amount required for two transmit
and receive antennas, respectively, which is a tremendous
reduction in the codeword length.

Table~\ref{table:2} shows the required codeword length $L$ for
i.i.d.\ and exponentially correlated ($\TxCC=0.5$, $\RxCC=0.7$)
MIMO channels with $\nTx=\nRx=3$ at $\snr=15$ dB when $\Tc$ varies
from $1$ to $10$. It is clear from Table~\ref{table:2} that for
each value of $\Tc$, the codeword lengths for correlated channels
are much longer than those for i.i.d.\ channels. For example, the
increase in the required codeword length, due to exponential
correlation ($\TxCC=0.5$, $\RxCC=0.7$), ranges from $194{\%}$ for
$\Tc=1$ to $138{\%}$ for $\Tc=10$, which is a significant increase
in required codeword length. Also, when going $\Tc$ from $1$ to
$10$, there is a considerable increase in the required codeword
length, relative to that for the single-symbol coherence time,
which ranges from $33{\%}$ to $344{\%}$ for the i.i.d.\ case and
from $28{\%}$ to $258{\%}$ for the correlated case, respectively.

Table~\ref{table:3} demonstrates the effect of correlation on the
required code length $L$, where $\nTx=\nRx=3$, $\TxCC=\RxCC=\zeta
$, $\Tc=5$, and $\snr=15$ dB. The table contains the corresponding
codeword lengths for $\zeta$ from $0$ to $0.9$. As seen from the
table, the required codeword length for the case of exponential
correlation $\zeta =0.7$ is equal to $4.5$ times as long as for
the i.i.d.\ channel ($\zeta =0$). Particularly, when $\zeta \geq
0.5$, there exists a large amount of increase in required codeword
length due to a stronger correlation. Also, since the ergodic
capacity is $7.36$ bits/symbol for $\TxCC=\RxCC=0.9$ at $\snr=15$
dB, it is impossible to achieve reliable communications at a rate
$8.0$ bits/symbol (regardless of the codeword length), when
$\TxCC=\RxCC =0.9$.

\begin{figure}[t]
    \centerline{\includegraphics[width=0.7\textwidth]{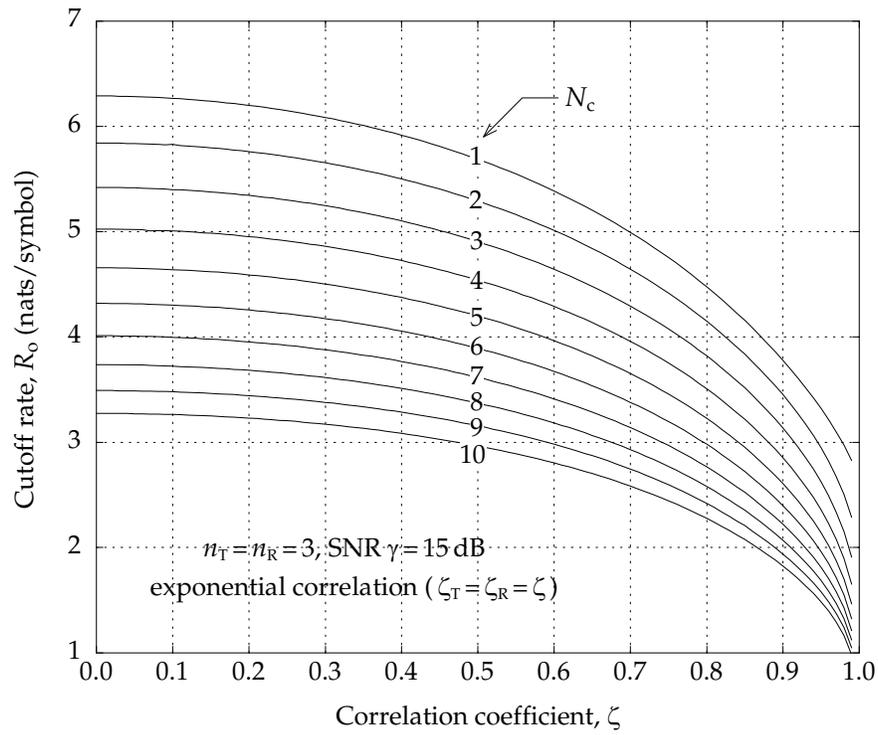}}
    \caption{
        Cutoff rate $R_0$ in nats/symbol as a function of a correlation coefficient $\zeta$
        for for exponentially correlated MIMO
        channels with $\TxCC=\RxCC=\zeta$ when $\Tc=1$, $2$, $3$,
        $4$, $5$, $6$, $7$, $8$, $9$, and $10$. $\nTx=\nRx=3$
        and $\snr=15$ dB.
    }
    \label{fig:5}
\end{figure}

Finally, Fig.~\ref{fig:5} shows the cutoff rate $R_0$ in
nats/symbol as a function of a correlation coefficient $\zeta$ for
exponentially correlated MIMO channels with $\TxCC=\RxCC=\zeta$ at
$\snr=15$ dB, where $\nRx=\nRx=3$ and $\Tc$ varies from $1$ to
$10$. We see that the cutoff rate $R_0$ decreases with $\Tc$ for
all $\zeta \in \left[0,1\right)$. While $\ErC$ remains constant,
$R_0$ monotonically decreases with $\Tc$, going to $0$ as $\Tc \to
\infty$ (see \eqref{eq:EoTc} and \eqref{eq:EoTITc} with $\rho=1$
and $\beta=\nTx$). Hence, these two measures diverge as $\Tc$
increases and eventually
%
%
$    \lim_{\Tc \to \infty}
        \frac{\ErC}{R_0}
    =
        \infty
$,
%
%
which coincides with the divergent behavior of the capacity and
cutoff rate of a channel with block memory \cite{MS:84:IT}. This
observation reveals that $R_0$ is more pertinent than $\ErC$ as a
figure of merit that reflects the quality of block-fading
channels.

\section{Conclusions}   \label{sec:Sec:5}

In this paper, we derived Gallager's random coding error exponent
to investigate a fundamental tradeoff between the communication
reliability and information rate in spatially correlated MIMO
channels. We considered a block-fading channel with perfect
receive CSI and Gaussian codebooks. The required codeword lengths
for a prescribed error probability were calculated from the random
coding bound to aid in the assessment of the coding requirement on
such MIMO channels, taking into account the effects of the number
of antennas, the channel coherence time, and the amount of spatial
fading correlation. In addition, we obtained the general formulae
for the ergodic capacity and cutoff rate, which encompass all the
previous capacity results as special cases of our expressions. In
parallel to the capacity--cutoff rate divergence in a block-memory
channel, we observed the channel-incurable effect: the
monotonically decreasing property of the MIMO exponent (i.e.,
communication reliability) with the channel coherence time.

\appendix

\subsection{Proof of Proposition~\ref{thm:EoGI}}    \label{sec:Appendix:I}

\begin{lemma}   \label{le:CGL}

Let $\B{S} \sim \CGM{m}{n}{\B{M}}{\B{\Sigma}}{\B{I}_n}$ and $\B{A}
\in \C^{m \times m}> 0$ be Hermitian. Then, we have
\begin{align}   \label{eq:leCGL}
\EqF
    \EX{
        \etr\left(-\B{A SS}^\dag\right)
       }
    =
        \det\left(
                    \B{I}_m
                    +
                    \B{\Sigma}
                    \B{A}
            \right)^{-n}
        \etr\left\{
                    -\left(
                            \B{A}^{-1}
                            +
                            \B{\Sigma}
                     \right)^{-1}
                     \B{MM}^\dag
            \right\}.
\EqE
\end{align}

\begin{proof}
Note that
\begin{align}   \label{eq:leCGLp1}
    \EX{
        \etr\left(-\B{A SS}^\dag\right)
       }
    =
        \frac{
                \det\left(\B{\Sigma}\right)^{-n}
             }
             {
                \pi^{mn}
             }
        \int_{\B{S}}
            \etr\left\{
                        -\B{A SS}^\dag
                        -\B{\Sigma}^{-1}
                         \left(\B{S}-\B{M}\right)
                         \left(\B{S}-\B{M}\right)^\dag
                \right\}
        d\B{S}.
\EqE
\end{align}
By writing the trace of the quadratic form in the exponent of
\eqref{eq:leCGLp1} as
\begin{align}   \label{eq:leCGLp2}
\EqF
    &
    \tr\left\{
                \B{A SS}^\dag
                +
                \B{\Sigma}^{-1}
                \left(\B{S}-\B{M}\right)
                \left(\B{S}-\B{M}\right)^\dag
        \right\}
    \nonumber \\
    & \quad
    =
        \tr\left\{
                    \left(\B{A}+\B{\Sigma}^{-1}\right)
                    \left[
                            \B{S}
                            -
                            \left(\B{I}_m + \B{\Sigma}\B{A}\right)^{-1}
                            \B{M}
                    \right]
                    \left[
                            \B{S}
                            -
                            \left(\B{I}_m + \B{\Sigma}\B{A}\right)^{-1}
                            \B{M}
                    \right]^\dag
                    -
                    \left(
                            \B{A}^{-1} + \B{\Sigma}
                    \right)^{-1}
                    \B{MM}^\dag
            \right\},
\end{align}
we get
\begin{align}   \label{eq:leCGLp3}
    &\EX{
        \etr\left(-\B{A SS}^\dag\right)
       }
    =
        \frac{
                \det\left(\B{\Sigma}\right)^{-n}
             }
             {
                \pi^{mn}
             }
        \etr\left\{
                    -\left(
                            \B{A}^{-1}
                            +
                            \B{\Sigma}
                     \right)^{-1}
                     \B{MM}^\dag
            \right\}
    \nonumber \\
    & \hspace{1cm}
    \times
        \underbrace{
        \int_{\B{S}}
            \etr\left\{
                        -
                        \left(\B{A}+\B{\Sigma}^{-1}\right)
                        \left[
                                \B{S}
                                -
                                \left(\B{I}_m + \B{\Sigma}\B{A}\right)^{-1}
                                \B{M}
                        \right]
                        \left[
                                \B{S}
                                -
                                \left(\B{I}_m + \B{\Sigma}\B{A}\right)^{-1}
                                \B{M}
                        \right]^\dag
                \right\}
           d\B{S}
        }_{
            =
            \pi^{mn}
            \det\left(
                        \B{A}
                        +
                        \B{\Sigma}^{-1}
                \right)^{-n}
           }
\end{align}
from which \eqref{eq:leCGL} follows readily.
\end{proof}

\end{lemma}


\textit{Proof of Proposition~\ref{thm:EoGI}:}
Using Lemma~\ref{le:CGL}, we have
\begin{align}   \label{eq:EoGId1}
\EqF
    &
    \int_{\B{X}}
        \PDF{\B{X}}{\B{X}}
        e^{
            r
            \left[
                    \tr\left(\B{XX}^\dag\right)
                    -\Tc\mathcal{P}
            \right]
        }
        p\left(
                \left. \B{Y} \right|
                \B{X},
                \B{H}
         \right)^{1/\left(1+\rho\right)}
        d\B{X}
    \nonumber \\
    & \quad
    =
        e^{
            -r
             \Tc
             \mathcal{P}
          }
        \left(\pi N_0\right)^{-\nRx \Tc/\left(1+\rho\right)}
        \det\left(
                    \B{I}_\nTx
                    -
                    r
                    \B{Q}
            \right)^{-\Tc}
        \det\left(
                    \B{I}_\nRx
                    +
                    \frac{
                            \B{H}
                            \left(
                                    \B{Q}^{-1}
                                    -
                                    r
                                    \B{I}_\nTx
                            \right)^{-1}
                            \B{H}^\dag
                         }
                         {
                            N_0
                            \left(1+\rho\right)
                         }
            \right)^{-\Tc}
    \nonumber \\
    & \hspace{1.5cm}
    \times
        \etr\left\{
                    -\frac{1}{N_0 \left(1+\rho\right)}
                     \left(
                            \B{I}_\nRx
                            +
                            \frac{
                                    \B{H}
                                    \left(
                                            \B{Q}^{-1}
                                            -
                                            r
                                            \B{I}_\nTx
                                    \right)^{-1}
                                    \B{H}^\dag
                                }
                                {
                                    N_0
                                    \left(1+\rho\right)
                                }
                     \right)^{-1}
                     \B{YY}^\dag
             \right\}.
\EqE
\end{align}
Substituting \eqref{eq:EoGId1} into \eqref{eq:Eo} and integrating
over $\B{Y}$, we have
\begin{align}   \label{eq:EoGId2}
\EqF
    &
    \int_{\B{Y}}
        \left\{
                \int_{\B{X}}
                    \PDF{\B{X}}{\B{X}}
                    e^{
                        r
                        \left[
                                \tr\left(\B{XX}^\dag\right)
                                -\Tc\mathcal{P}
                        \right]
                    }
                    p\left(
                            \left. \B{Y} \right|
                            \B{X},
                            \B{H}
                     \right)^{1/\left(1+\rho\right)}
                    d\B{X}
        \right\}^{1+\rho}
        d\B{Y}
    \nonumber \\
    & \quad
    =
        e^{
            -r
             \Tc
             \mathcal{P}
             \left(1+\rho\right)
        }
        \det\left(
                    \B{I}_\nTx
                    -
                    r
                    \B{Q}
            \right)^{-\Tc\left(1+\rho\right)}
        \det\left(
                    \B{I}_\nRx
                    +
                    \frac{
                            \B{H}
                            \left(
                                    \B{Q}^{-1}
                                    -
                                    r
                                    \B{I}_\nTx
                            \right)^{-1}
                            \B{H}^\dag
                         }
                         {
                            N_0
                            \left(1+\rho\right)
                         }
            \right)^{-\Tc\rho}.
\EqE
\end{align}
Finally, substituting \eqref{eq:EoGId2} into \eqref{eq:Eo} yields
the result \eqref{eq:EoGI}.

\subsection{Proof of Proposition \ref{thm:OB}} \label{sec:Appendix:II}

We provide a sketch of the proof of Proposition \ref{thm:OB} using
a similar approach in \cite{Eri:70:IT} and \cite{AM:99:IT}. For
notational simplicity, let us denote $\B{\Omega}_{\rho,\beta}=
\beta \B{I}_\nXmin + \snr \MHH / \left(1+\rho\right)$. Then,
$\EoT{\rho}{\beta}{\Tc}$ in \eqref{eq:EoT} can be rewritten as
\begin{align}   \label{eq:EoTA}
    \EoT{\rho}{\beta}{\Tc}
    &=
        \EoConst{\rho}{\beta}
        -
        \nXmin
        \rho
        \ln\left(\beta\right)
        -
        \frac{1}{\Tc}
        \ln
        \EoRFA{\rho}{\beta}{\Tc}{}
\end{align}
where
%
%
%
%
$   \EoRFA{\rho}{\beta}{\Tc}{}
    =
        \E\bigl\{
            \det\left(
                \B{\Omega}_{\rho,\beta}
                \right)^{-\Tc\rho}
           \bigr\}$.
Since $\EoConst{\rho}{\beta}-\nXmin\rho \ln\left(\beta\right)$ is
concave in $\beta$, $\EoT{\rho}{\beta}{\Tc}$ is a concave function
of $\beta$ if $\ln\EoRFA{\rho}{\beta}{\Tc}{-1}$ is concave in
$\beta$ for all $\rho \in \left[0,1\right]$. Noting that
\begin{align}   \label{eq:ZoIDD}
\EqF
    \frac{
            \partial^2
            \ln
            \EoRFA{\rho}{\beta}{\Tc}{-1}
         }
         {
            \partial
            \beta^2
         }
    =
        \EoRFA{\rho}{\beta}{\Tc}{-2}
        \left\{
                \left(
                        \frac{
                                \partial
                                \EoRFA{\rho}{\beta}{\Tc}{}
                             }
                             {
                                \partial
                                \beta
                             }
                \right)^2
                -
                \EoRFA{\rho}{\beta}{\Tc}{}
                \frac{
                        \partial^2
                        \EoRFA{\rho}{\beta}{\Tc}{}
                     }
                     {
                        \partial
                        \beta^2
                     }
        \right\}
\end{align}
and $\EoRFA{\rho}{\beta}{\Tc}{} \geq 0$, it is sufficient to show
that
\begin{align}   \label{eq:ZoDB}
\EqF
    \left(
            \frac{
                    \partial
                    \EoRFA{\rho}{\beta}{\Tc}{}
                 }
                 {
                    \partial
                    \beta
                 }
    \right)^2
    \leq
        \EoRFA{\rho}{\beta}{\Tc}{}
        \frac{
                \partial^2
                \EoRFA{\rho}{\beta}{\Tc}{}
             }
             {
                \partial
                \beta^2
             } \,.
\EqE
\end{align}

It is easy to show that
%
%
%
%
\begin{align*}
\EqF
    \frac{
            \partial
            \det\left(
                        \B{\Omega}_{\rho,\beta}
                \right)
         }
         {
            \partial
            \beta
         }
    =
        \det\left(
                    \B{\Omega}_{\rho,\beta}
            \right)
        \,
        \tr\left(
                    \B{\Omega}_{\rho,\beta}^{-1}
            \right)
    \quad
    \text{and}
    \quad
    \frac{
            \partial
            \tr\left(
                        \B{\Omega}_{\rho,\beta}^{-1}
                \right)
         }
         {
            \partial
            \beta
         }
    =
        -\tr\left(
                    \B{\Omega}_{\rho,\beta}^{-2}
            \right)
\end{align*}
and hence,
\begin{align}   \label{eq:DZo}
    \frac{
            \partial
            \EoRFA{\rho}{\beta}{\Tc}{}
         }
         {
            \partial
            \beta
         }
    &=
        \EX{
            -\Tc
             \rho
             \,
             \det\left(
                        \B{\Omega}_{\rho,\beta}
                 \right)^{-\Tc\rho}
             \,
             \tr\left(
                        \B{\Omega}_{\rho,\beta}^{-1}
                \right)
          }
    \\
    \label{eq:DDZo}
    \frac{
            \partial^2
            \EoRFA{\rho}{\beta}{\Tc}{}
         }
         {
            \partial
            \beta^{2}
         }
    &=
        \EX{
            \Tc
            \rho
            \,
            \det\left(
                        \B{\Omega}_{\rho,\beta}
                 \right)^{-\Tc\rho}
            \left[
                    \Tc
                    \rho
                    \,
                    \tr^2\left(
                                \B{\Omega}_{\rho,\beta}^{-1}
                        \right)
                    +
                    \tr\left(
                                \B{\Omega}_{\rho,\beta}^{-2}
                        \right)
            \right]
           }.
\EqE
\end{align}

Let us now define the random variables
%
%
%
%
\begin{align*}
    \mathsf{X}^2
    =
        \det\left(
                    \B{\Omega}_{\rho,\beta}
            \right)^{-\Tc\rho}
    \quad
    \text{and}
    \quad
    \mathsf{Y}^2
    =
        \left(\Tc\rho\right)^2
        \,
        \det\left(
                    \B{\Omega}_{\rho,\beta}
            \right)^{-\Tc\rho}
        \,
        \tr^2\left(
                    \B{\Omega}_{\rho,\beta}^{-1}
            \right).
\EqE
\end{align*}
From Schwartz's inequality, we have
\begin{align}   \label{eq:DDZoS}
\EqF
    \left(
            \frac{
                    \partial
                    \EoRFA{\rho}{\beta}{\Tc}{}
                 }
                 {
                    \partial
                    \beta
                 }
    \right)^2
    &=
        \E^2\left\{\mathsf{XY}\right\}
    \leq
        \EX{\mathsf{X}^2}
        \cdot
        \EX{\mathsf{Y}^2}
    \nonumber \\
    &
    \leq
        \EX{\mathsf{X}^2}
        \cdot
        \EX{
            \mathsf{Y}^2
            +
            \Tc\rho \,
            \mathsf{X}^2 \,
            \tr\left(
                        \B{\Omega}_{\rho,\beta}^{-2}
                \right)
          }
    \nonumber \\
    &=
        \EoRFA{\rho}{\beta}{\Tc}{}
        \frac{
                \partial^2
                \EoRFA{\rho}{\beta}{\Tc}{}
             }
             {
                \partial
                \beta^2
             } \, .
\EqE
\end{align}
From \eqref{eq:DDZoS}, we see that $\EoT{\rho}{\beta}{\Tc}$ is a
concave function of $\beta$ for all $\rho \in \left[0,1\right]$.
Hence, the maximum over $\beta$ occurs at $\OB{\rho}$ for which $
\bigl[
\partial \EoT{\rho}{\beta}{\Tc}/
\partial \beta \bigr] \bigr|_{\beta=\OB{\rho}}=0$
%
%
%
%
and it is sufficient to show that $\bigl[\partial
\EoT{\rho}{\beta}{\Tc}/\partial\beta \bigr] \bigr|_{\beta=0} \geq
0$ and $\bigl[ \partial \EoT{\rho}{\beta}{\Tc}/\partial\beta
\bigr] \bigr|_{\beta=\nTx} \leq 0$
%
%
%
%
for all $\rho \in \left[0,1\right]$ in order to prove
$0<\OB{\rho}\leq \nTx$. Since
\begin{align}   \label{eq:DBEoT}
\EqF
    \frac{
            \partial
            \tilde{E}_0\left(\rho,\beta,\Tc\right)
         }
         {
            \partial
            \beta
         }
    =
        \underbrace{
            \frac{
                    \left(1+\rho\right)
                    \left(\nTx-\beta\right)
                 }
                 {\beta}
        }_{
            \triangleq \,
             \dBEoConst{\rho}{\beta}
             =
                \frac{
                        \partial
                        \EoConst{\rho}{\beta}
                     }
                     {
                        \partial
                        \beta
                     }
          }
        -
        \frac{
                \nXmin
                \rho
             }
             {\beta}
        -
        \frac{1}{\Tc}
        \EoRFA{\rho}{\beta}{\Tc}{-1}
        \frac{
                \partial
                \EoRFA{\rho}{\beta}{\Tc}{}
             }
             {
                \partial
                \beta
             },
\EqE
\end{align}
it is clear that $\lim_{\beta \rightarrow 0}
\partial \EoT{\rho}{\beta}{\Tc} /\partial\beta \rightarrow \infty$.
Also,
\begin{align}   \label{eq:DBEoTatNt}
\EqF
    \left.
    \left[
            \frac{
                    \partial
                    \EoT{\rho}{\beta}{\Tc}
                 }
                 {
                    \partial
                    \beta
                 }
    \right]
    \right|_{\beta=\nTx}
    =
        -\frac{\nXmin\rho}{\nTx}
        -\frac{1}{\Tc}
         \EoRFA{\rho}{\nTx}{\Tc}{-1}
         \left.
         \left[
                \frac{
                        \partial
                        \EoRFA{\rho}{\beta}{\Tc}{}
                     }
                     {
                        \partial
                        \beta
                     }
         \right]
         \right|_{\beta=\nTx}.
\EqE
\end{align}
Since $\EoRFA{\rho}{\beta}{\Tc}{} \geq 0$ and
\begin{align}   \label{eq:ZoatNt}
    \EoRFA{\rho}{\nTx}{\Tc}{}
    &=
        \EX{
            \det\left(
                        \B{\Omega}_{\rho,\nTx}
                \right)^{-\Tc\rho}
           }
    =
        \EX{
            \det\left(
                        \B{\Omega}_{\rho,\nTx}
                \right)^{-\Tc\rho}
            \,
            \frac{
                    \tr\left(
                                \B{\Omega}_{\rho,\nTx}^{-1}
                        \right)
                 }
                 {
                    \tr\left(
                                \B{\Omega}_{\rho,\nTx}^{-1}
                        \right)
                 }
            }
    \nonumber \\
    &
    \geq
        \frac{\nTx}{\nXmin}
        \EX{
            \det\left(
                        \B{\Omega}_{\rho,\nTx}
                \right)^{-\Tc\rho}
            \,
            \tr\left(
                        \B{\Omega}_{\rho,\nTx}^{-1}
                \right)
           }
    \nonumber \\
    &=
        -\frac{\nTx}{\nXmin\rho}
         \cdot
         \frac{1}{\Tc}
         \left.
         \left[
                \frac{
                        \partial
                        \EoRFA{\rho}{\beta}{\Tc}{}
                     }
                     {
                        \partial
                        \beta
                     }
         \right]
         \right|_{\beta=\nTx},
\end{align}
it follows that
\begin{align}   \label{eq:Pf}
    -\frac{\nXmin\rho}{\nTx}
    -\frac{1}{\Tc}
     \EoRFA{\rho}{\nTx}{\Tc}{-1}
     \left.
     \left[
            \frac{
                    \partial
                    \EoRFA{\rho}{\beta}{\Tc}{}
                 }
                 {
                    \partial
                    \beta
                 }
     \right]
     \right|_{\beta=\nTx}
     \leq 0.
\EqE
\end{align}
Thus, $\bigl[\partial \EoT{\rho}{\beta}{\Tc} /\partial\beta \bigr]
\bigr|_{\beta=\nTx} \leq 0$ and we complete the proof of the
proposition.


\clearpage

\begin{table}[t]
    \caption{
                Some quantities and matrices involved in
                Theorem~\ref{thm:RCEE}
    }
    \label{table:M}
\begin{center}
    \begin{tabular}{l}
        \hline
        \\[-0.5cm]
        \textbf{In Theorem~\ref{thm:RCEE}}
        \\[-0.5cm]
        \\ \hline \hline
        \\[-0.3cm]
            $
              \CM{\nu}{\B{\Psi}}
              =
                    \begin{bmatrix}
                        \pmb{\mathcal{A}}_1
                        &
                        \pmb{\mathcal{A}}_2
                        &
                        \cdots
                        &
                        \pmb{\mathcal{A}}_\EVn{\B{\Psi}}
                    \end{bmatrix}
                    \in \mathbb{R}^{\nu \times p},
              \qquad
              \text{for $\B{\Psi}$ $p \times p$ Hermitian, $~\nu \leq p$}
            $
        \\[0.1cm]
            $
              \overline{\B{\mathsf{G}}}_{(\nu)}\left(\B{\Psi}\right)
              =
                    \begin{bmatrix}
                        \overline{\pmb{\mathcal{A}}}_1
                        &
                        \overline{\pmb{\mathcal{A}}}_2
                        &
                        \cdots
                        &
                        \overline{\pmb{\mathcal{A}}}_\EVn{\B{\Psi}}
                    \end{bmatrix}
                    \in \mathbb{R}^{\nu \times p}
            $
        \\
            $\hspace{0.5cm}$
            where$^\text{1)}$
        \\
            $\hspace{1cm}$

            $\pmb{\mathcal{A}}_k =\left(\mathcal{A}_{k,ij}\right) \in \mathbb{R}^{\nu \times
            \EVm{k}{\B{\Psi}}}$, $k=1,2,\ldots,\EVn{\B{\Psi}}$
        \\
            $\hspace{2cm}$

            $\left(i,j\right)^{\underline{\text{th}}}$ element:
            $
                ~~
                \mathcal{A}_{k,ij}
                =
                    \left(-1\right)^{i-j}
                    \PS{i-j+1}{j-1}
                    \oEV{k}{-i+j}{\B{\Psi}}
            $
        \\[0.2cm]
            $\hspace{1cm}$

            $\overline{\pmb{\mathcal{A}}}_k =\left(\overline{\mathcal{A}}_{k,ij}\right) \in \mathbb{R}^{\nu \times
            \EVm{k}{\B{\Psi}}}$, $k=1,2,\ldots,\EVn{\B{\Psi}}$
        \\
            $\hspace{2cm}$

            $\left(i,j\right)^{\underline{\text{th}}}$ element:
            $
                ~~
                \overline{\mathcal{A}}_{k,ij}
                =
                    \left(-1\right)^{i-j}
                    \PS{i-j+1}{j-1}
                    \oEV{k}{i-j}{\B{\Psi}}.
            $
        \\[0.2cm]
            $^\text{1)} \PS{a}{n}=a\left(a+1\right)\cdots\left(a+n-1\right)$,
            $\PS{a}{0}=1$ is the Pochhammer symbol.
        \\[-0.3cm]
        \\ \hline
        \\[-0.3cm]
            $
              \Kc
              =
                    \det\left(
                            \CMmin
                        \right)^{\Tc\rho}
                    \det\left\{
                            \CM{\nXmin}{\CMmin}
                        \right\}
                    \prod\limits_{k=1}^{\Tc\rho}
                        \left(k-1\right)!
              \, ,
              \qquad
              \Tc\rho \in \left\{1,2,\ldots,\nXmin\right\}
            $
        \\[-0.3cm]
        \\ \hline
        \\[-0.3cm]
            $
              \Kiid
              =
                    \prod\limits_{k=1}^{\nXmin}
                        \left(\nXmax-k\right)!
                        \left(k-1\right)!
            $
        \\[-0.3cm]
        \\ \hline
        \\[-0.3cm]
            $
                \mathcal{T}_\text{A}
                =
                    \frac{1}{2}
                    \nXmin
                    \left(\nXmin+1\right)
                    -
                    \frac{1}{2}
                    \sum\limits_{i=1}^\EVn{\CMmin}
                        \EVm{i}{\CMmin}
                        \Bigl[\EVm{i}{\CMmin}+1\Bigr]
            $
        \\[-0.3cm]
        \\ \hline
        \\[-0.3cm]
            $
                \mathcal{T}_\text{B}\left(\rho,\Tc\right)
                =
                    \det\left(
                            \CMmax
                        \right)^{-\nXmin}
                    \det\bigl\{
                            \overline{\B{\mathsf{G}}}_{(\nXmin)}\left(\CMmin\right)
                        \bigr\}^{-1}
                    \det\left\{
                            \CM{\nXmax}{\CMmax}
                        \right\}^{-1}
                    \dfrac{
                            \prod\nolimits_{i=1}^\EVn{\CMmin}
                            \prod\nolimits_{j=1}^\EVm{i}{\CMmin}
                            \PS{\Tc\rho-\nXmin+1}{j-1}
                        }{
                            \prod\nolimits_{k=1}^\nXmin
                            \PS{\Tc\rho-\nXmin+1}{k-1}
                        }
            $
        \\[-0.3cm]
        \\ \hline
        \\[-0.3cm]
            $
                \B{\Xi}\left(\rho,\beta\right)
                =
                    \begin{bmatrix}
                        \B{\Xi}_1\left(\rho,\beta\right)
                        &
                        \B{\Xi}_2\left(\rho,\beta\right)
                        &
                        \cdots
                        &
                        \B{\Xi}_{\EVn{\CMmin}}\left(\rho,\beta\right)
                    \end{bmatrix}
                    \in \mathbb{R}^{\Tc\rho \times \nXmin}
            $
        \\
            $\hspace{0.5cm}$
            where$^\text{2)}$
        \\
            $\hspace{1cm}$

            $\B{\Xi}_{k}\left(\rho,\beta\right)=\left(\Xi_{k,ij}\left(\rho,\beta\right)\right)
            \in \mathbb{R}^{\Tc\rho \times \EVm{k}{\CMmin}}$,
            $k=1,2,\ldots,\EVn{\CMmin}$,
            $\Tc\rho \in \left\{1,2,\ldots,\nXmin\right\}$
        \\
            $\hspace{2cm}$

            $\left(i,j\right)^{\underline{\text{th}}}$ element:
        \\
            $\hspace{2cm}$

            $
                \Xi_{k,ij}\left(\rho,\beta\right)
                =
                    \sum\nolimits_{p=1}^{\EVn{\CMmax}}
                    \sum\nolimits_{q=1}^{\EVm{p}{\CMmax}}
                        \PFC{p}{q}{\CMmax}
                        \mathcal{G}_{i+j-1,1}
                            \left(
                                    \frac{\snr}{\beta\left(1+\rho\right)}
                                    \oEV{p}{}{\CMmax},
                                    \oEV{k}{}{\CMmin},
                                    -q+1
                            \right)
            $.

        \\[0.3cm]
            $^\text{2)}$
            $\PFC{p}{q}{\CMmax}$ is the $\left(p,q\right)$th \emph{characteristic coefficient}
            of $\CMmax$ (see for details \cite[Definition~6]{SW:05:IT}).
        \\
            $~~\,$The $\mathcal{G}_{\kappa,\nu} \left(a,b,\mu\right)$ is defined
            as the integral
        \\[0.2cm]
            $\hspace{1cm}$

            $
                \mathcal{G}_{\kappa,\nu} \left(a,b,\mu\right)
                =
                    \int\limits_0^\infty
                        \left(1+ax\right)^{\mu-1}
                        \ln^{\nu-1} \left(1+ax\right)
                        x^{\kappa-1}
                        e^{-x/b}
                    dx
            $,
            $\quad$
            $a,b>0$, $\kappa,\nu \in \mathbb{\Nb}$, $\mu \in \C$
        \\
            $\hspace{1.2cm}$

            $
                =
                \begin{cases}
                        b^{\kappa}
                        \left(\kappa-1\right)! \;
                        \HyperPFQ{2}{0}{\kappa,-\mu+1;-ab},
                    &
                        \text{if $\nu=1$}
                    \\
                        a^{-\kappa}
                        \left(\nu-1\right)! \;
                        e^{1/\left(ab\right)}
                        \sum\limits_{k=0}^{\kappa-1}
                            \Bigl[
                                \left(-1\right)^{\kappa-k-1}
                                \binom{\kappa-1}{k}
                                \left(ab\right)^{\mu+k}
                                G_{\nu,\nu+1}^{\nu+1,0}
                                    \Bigl(
                                            \frac{1}{ab}
                                            \Bigr|
                                            \begin{subarray}{l}
                                                1,1,\ldots,1
                                                \\
                                                0,0,\ldots,0,
                                                \mu+k
                                            \end{subarray}
                                    \Bigr)
                            \Bigr],
                    &
                        \text{otherwise}
                \end{cases}
            $

        \\[0.2cm]
            where $\HyperPFQ{p}{q}{a_1,a_2,\ldots,a_p;b_1,b_2,\ldots,b_q;z}$
            is the generalized hypergeometric function of scalar argument
        \\
            \cite[eq. (9.14.1)]{GR:00:Book} and $G_{p,q}^{m,n}
            \left(\cdot\right)$ is the Meijer G-function
            \cite[eq.~(9.301)]{GR:00:Book}. The detailed derivation of
            this
        \\
            integral identity can be found in \cite[Appendix~A]{SWLC:05:WCOM}.

        \\[-0.3cm]
        \\ \hline
    \end{tabular}
\end{center}
\end{table}

\clearpage

\addtocounter{table}{-1}

\begin{table}[t]
    \caption{
                \textit{(Continued.)}
                Some quantities and matrices involved in
                Theorem~\ref{thm:RCEE}
    }
    \label{table:M}
\begin{center}
    \begin{tabular}{l}
        \hline
        \\[-0.5cm]
        \textbf{In Theorem~\ref{thm:RCEE}}
        \\[-0.5cm]
        \\ \hline \hline
        \\[-0.3cm]
            $
                \B{\Upsilon}\left(\rho,\beta\right)
                =
                    \begin{bmatrix}
                        \B{\Upsilon}_{1,1}\left(\rho,\beta\right)
                        &
                        \cdots
                        &
                        \B{\Upsilon}_{1,\EVn{\CMmax}}\left(\rho,\beta\right)
                        \\
                        \vdots
                        &
                        \ddots
                        &
                        \vdots
                        \\
                        \B{\Upsilon}_{\EVn{\CMmin},1}\left(\rho,\beta\right)
                        &
                        \cdots
                        &
                        \B{\Upsilon}_{\EVn{\CMmin},\EVn{\CMmax}}\left(\rho,\beta\right)
                    \end{bmatrix}
                    \in \mathbb{R}^{\nXmin \times \nXmax}
            $
        \\
            $\hspace{0.5cm}$
            where
        \\
            $\hspace{1cm}$

            $\B{\Upsilon}_{p,q}\left(\rho,\beta\right)=\left(\Upsilon_{p,q,ij}\left(\rho,\beta\right)\right)
            \in \mathbb{R}^{\EVm{p}{\CMmin} \times
            \EVm{q}{\CMmax}}$,
            $p=1,2,\ldots,\EVn{\CMmin}$,
            $q=1,2,\ldots,\EVn{\CMmax}$
        \\
            $\hspace{2cm}$

            $\left(i,j\right)^{\underline{\text{th}}}$ element:

            $
                ~~
                \Upsilon_{p,q,ij}\left(\rho,\beta\right)
                =
                    \mathcal{G}_{i+j-1,1}
                        \left(
                                \frac{\snr}{\beta\left(1+\rho\right)}
                                \oEV{p}{}{\CMmin},
                                \oEV{q}{}{\CMmax},
                                -\Tc\rho+\nXmin-i+1
                        \right)
            $.

        \\[-0.3cm]
        \\ \hline
        \\[-0.3cm]
            $
                \B{\Upsilon}^{\left(\beta\right)}\left(\rho,\beta\right)
                \triangleq
                    \frac{
                            \partial
                    }{
                            \partial
                            \beta
                    }
                    \B{\Upsilon}\left(\rho,\beta\right)
                =
                    \begin{bmatrix}
                        \B{\Upsilon}^{\left(\beta\right)}_{1,1}\left(\rho,\beta\right)
                        &
                        \cdots
                        &
                        \B{\Upsilon}^{\left(\beta\right)}_{1,\EVn{\CMmax}}\left(\rho,\beta\right)
                        \\
                        \vdots
                        &
                        \ddots
                        &
                        \vdots
                        \\
                        \B{\Upsilon}^{\left(\beta\right)}_{\EVn{\CMmin},1}\left(\rho,\beta\right)
                        &
                        \cdots
                        &
                        \B{\Upsilon}^{\left(\beta\right)}_{\EVn{\CMmin},\EVn{\CMmax}}\left(\rho,\beta\right)
                    \end{bmatrix}
            $
        \\
            $\hspace{0.5cm}$
            where
        \\
            $\hspace{1cm}$

            $
                \B{\Upsilon}^{\left(\beta\right)}_{p,q}\left(\rho,\beta\right)
                =
                    \bigl(
                            \B{\Upsilon}^{\left(\beta\right)}_{p,q,ij}\left(\rho,\beta\right)
                    \bigr)
                    \in \mathbb{R}^{\EVm{p}{\CMmin} \times \EVm{q}{\CMmax}}
            $
        \\
            $\hspace{2cm}$

            $\left(i,j\right)^{\underline{\text{th}}}$ element:
        \\
            $\hspace{2cm}$

            $
                \Upsilon_{p,q,ij}^{\left(\beta\right)}\left(\rho,\beta\right)
                =
                    \frac{\snr}{\beta^2\left(1+\rho\right)}
                    \left(
                        \Tc\rho-\nXmin+i
                    \right)
                    \oEV{p}{}{\CMmin}
                    \,
                    \mathcal{G}_{i+j,1}
                        \left(
                            \frac{\snr}{\beta\left(1+\rho\right)}
                            \oEV{p}{}{\CMmin},
                            \oEV{q}{}{\CMmax},
                            -\Tc\rho+\nXmin-i
                        \right)
            $.
        \\[-0.3cm]
        \\ \hline
        \\[-0.3cm]
            $
                \B{\Upsilon}^{\left(\rho\right)}\left(\rho,\beta\right)
                \triangleq
                    \frac{
                            \partial
                    }{
                            \partial
                            \rho
                    }
                    \B{\Upsilon}\left(\rho,\beta\right)
                =
                    \begin{bmatrix}
                        \B{\Upsilon}^{\left(\rho\right)}_{1,1}\left(\rho,\beta\right)
                        &
                        \cdots
                        &
                        \B{\Upsilon}^{\left(\rho\right)}_{1,\EVn{\CMmax}}\left(\rho,\beta\right)
                        \\
                        \vdots
                        &
                        \ddots
                        &
                        \vdots
                        \\
                        \B{\Upsilon}^{\left(\rho\right)}_{\EVn{\CMmin},1}\left(\rho,\beta\right)
                        &
                        \cdots
                        &
                        \B{\Upsilon}^{\left(\rho\right)}_{\EVn{\CMmin},\EVn{\CMmax}}\left(\rho,\beta\right)
                    \end{bmatrix}
            $
        \\
            $\hspace{0.5cm}$
            where
        \\
            $\hspace{1cm}$

            $
                \B{\Upsilon}^{\left(\rho\right)}_{p,q}\left(\rho,\beta\right)
                =
                    \bigl(
                            \B{\Upsilon}^{\left(\rho\right)}_{p,q,ij}\left(\rho,\beta\right)
                    \bigr)
                    \in \mathbb{R}^{\EVm{p}{\CMmin} \times \EVm{q}{\CMmax}}
            $
        \\
            $\hspace{2cm}$

            $\left(i,j\right)^{\underline{\text{th}}}$ element:
        \\
            $\hspace{2cm}$

            $
                \Upsilon_{p,q,ij}^{\left(\rho\right)}\left(\rho,\beta\right)
                =
                    \frac{\beta}{1+\rho}
                    \Upsilon_{p,q,ij}^{\left(\beta\right)}\left(\rho,\beta\right)
                    -
                    \Tc \,
                    \mathcal{G}_{i+j-1,2}
                        \left(
                                \frac{\snr}{\beta\left(1+\rho\right)}
                                \oEV{p}{}{\CMmin},
                                \oEV{q}{}{\CMmax},
                                -\Tc\rho+\nXmin-i+1
                    \right)
            $.
        \\[-0.3cm]

        \\ \hline
        \\[-0.3cm]
            $
                \B{\Upsilon}_\text{iid}\left(\rho,\beta\right)
                =
                    \left(\Upsilon_{\text{iid},ij}\left(\rho,\beta\right)\right)
                    \in \mathbb{R}^{\nXmin \times \nXmin}
            $
        \\
            $\hspace{1cm}$

            $\left(i,j\right)^{\underline{\text{th}}}$ element:

            $
                \Upsilon_{\text{iid},ij}\left(\rho,\beta\right)
                =
                    \mathcal{G}_{\nXmax-\nXmin+i+j-1,1}
                        \left(
                                \frac{\snr}{\beta\left(1+\rho\right)},
                                1,
                                -\Tc\rho+1
                        \right)
            $
        \\[-0.3cm]
        \\ \hline
        \\[-0.3cm]
            $
                \B{\Upsilon}_\text{iid}^{\left(\beta\right)}\left(\rho,\beta\right)
                \triangleq
                    \frac{
                            \partial
                    }{
                            \partial
                            \beta
                    }
                    \B{\Upsilon}_\text{iid}\left(\rho,\beta\right)
                =
                    \bigl(
                            \Upsilon^{\left(\beta\right)}_{\text{iid},ij}\left(\rho,\beta\right)
                    \bigr)
                    \in \mathbb{R}^{\nXmin \times \nXmin}
            $
        \\
            $\hspace{1cm}$

            $\left(i,j\right)^{\underline{\text{th}}}$ element:

            $
                \Upsilon^{\left(\beta\right)}_{\text{iid},ij}\left(\rho,\beta\right)
                =
                    \frac{
                            \Tc
                            \rho
                            \snr
                    }{
                            \beta^2\left(1+\rho\right)
                    }
                    \,
                    \mathcal{G}_{\nXmax-\nXmin+i+j,1}
                        \left(
                                \frac{\snr}{\beta\left(1+\rho\right)},
                                1,
                                -\Tc\rho
                        \right)
            $
        \\[-0.3cm]
        \\ \hline
        \\[-0.3cm]
            $
                \B{\Upsilon}_\text{iid}^{\left(\rho\right)}\left(\rho,\beta\right)
                \triangleq
                    \frac{
                            \partial
                    }{
                            \partial
                            \rho
                    }
                    \B{\Upsilon}_\text{iid}\left(\rho,\beta\right)
                =
                    \bigl(
                            \Upsilon^{\left(\rho\right)}_{\text{iid},ij}\left(\rho,\beta\right)
                    \bigr)
                    \in \mathbb{R}^{\nXmin \times \nXmin}
            $
        \\
            $\hspace{1cm}$

            $\left(i,j\right)^{\underline{\text{th}}}$ element:

            $
                \Upsilon^{\left(\rho\right)}_{\text{iid},ij}\left(\rho,\beta\right)
                =
                    \frac{\beta}{1+\rho}
                    \Upsilon^{\left(\beta\right)}_{\text{iid},ij}\left(\rho,\beta\right)
                    -
                    \Tc \,
                    \mathcal{G}_{\nXmax-\nXmin+i+j-1,2}
                        \left(
                                \frac{\snr}{\beta\left(1+\rho\right)},
                                1,
                                -\Tc\rho+1
                        \right)
            $
        \\[-0.3cm]
        \\ \hline

    \end{tabular}
\end{center}
\end{table}

\clearpage

\begin{table}[t]
    \caption{
                Required codeword length $L$ as a function of SNR $\snr$ for
                i.i.d.\ MIMO channels ($\TxCM=\B{I}_\nTx$, $\RxCM=\B{I}_\nRx$)
                at a rate $8.0$ bits/symbol with $P_e \leq 10^{-6}$
                for different numbers of antennas and
                $\Tc=5$
    }
    \label{table:1}
    \begin{center}

    \begin{tabular}{c|c|c|c}
        \hline
        \raisebox{-1.70ex}[0cm][0cm]{$~~~\text{SNR (dB)}~~~$}
        &
            \multicolumn{3}{c}{Codeword length $L$}
            \\ \cline{2-4}
            & $\nTx=\nRx=2$
            & $\nTx=\nRx=3$
            & $\nTx=\nRx=4$
        \\ \hline \hline
        $8$ & - & - & $510$
        \\
        $10$ & - & $10865$ & $75$
        \\
        $12$ & - & $210$ & $30$
        \\
        $14$ & - & $65$ & $15$
        \\
        $16$ & $1070$ & $30$ & $10$
        \\
        $18$ & $205$ & $20$ & $5$
        \\
        $20$ & $90$ & $15$ & $5$
        \\ \hline

        \multicolumn{4}{l}{
            \begin{minipage}{7cm}

                $\begin{array}{l}
                    \\[-0.4cm]
                    \textit{Note: }
                    \text{The ergodic capacity $\ErC$ of $8.0$ bits/symbol is attained at}
                    \\[-0.15cm]
                    \text{$\snr=14.55$ dB for $\nTx=\nRx=2$; $\snr=9.68$ dB for $\nTx=\nRx=3$;}
                    \\[-0.15cm]
                    \text{and $\snr=6.79$ dB for $\nTx=\nRx=4$, respectively.}
                \end{array}$
            \end{minipage}
        }

    \end{tabular}
    \end{center}
\end{table}


\begin{table}[t]
    \caption{
                Required codeword length $L$ as a function of channel
                coherence time $\Tc$ for i.i.d.\ and exponentially correlated MIMO
                channels at a rate $8.0$ bits/symbol for $P_e \leq
                10^{-6}$, $\nTx=\nRx=3$, and $\snr=15$ $\mathrm{dB}$
    }
    \label{table:2}
\begin{center}
    \begin{tabular}{c|c|c}
        \hline
        \raisebox{-1.70ex}[0cm][0cm]
            {
                    Coherence time
                    $\Tc$
            }
        &
            \multicolumn{2}{c}{Codeword length $L$}
            \\ \cline{2-3}
            &
                \begin{minipage}{3cm}
                    \centering
                    i.i.d.
                \end{minipage}
            &
                \begin{minipage}{3cm}
                    \centering
                    $\TxCC=0.5$, $\RxCC=0.7$
                \end{minipage}
        \\ \hline \hline
        $1$ & $18$ & $53$
        \\
        $2$ & $24$ & $68$
        \\
        $3$ & $30$ & $84$
        \\
        $4$ & $36$ & $100$
        \\
        $5$ & $45$ & $115$
        \\
        $6$ & $48$ & $126$
        \\
        $7$ & $56$ & $140$
        \\
        $8$ & $64$ & $160$
        \\
        $9$ & $72$ & $171$
        \\
        $10$ & $80$ & $190$
        \\ \hline
    \end{tabular}
\end{center}
\end{table}

\clearpage

\begin{table}[t]
    \caption{
                Required codeword length $L$ as a function of correlation
                coefficient $\zeta$ for exponentially correlated MIMO channels
                with $\TxCC=\RxCC=\zeta$ at a rate $8.0$ bits/symbol for
                $P_e \leq 10^{-6}$, $\nTx=\nRx=$3, $\Tc=5$, and $\snr=15$ $\mathrm{dB}$
    }
    \label{table:3}
\begin{center}
    \begin{tabular}{c|c}
        \hline
            \begin{minipage}{3.5cm}
                \centering
                Correlation coefficient $\zeta$
            \end{minipage}
            &
            \begin{minipage}{3.5cm}
                \centering
                Codeword length $L$
            \end{minipage}
        \\ \hline \hline
        $0.0$ & $45$
        \\
        $0.1$ & $45$
        \\
        $0.2$ & $45$
        \\
        $0.3$ & $50$
        \\
        $0.4$ & $60$
        \\
        $0.5$ & $75$
        \\
        $0.6$ & $105$
        \\
        $0.7$ & $200$
        \\
        $0.8$ & $1015$
        \\
        $0.9$ & -
        \\ \hline

        \multicolumn{2}{l}{
            \begin{minipage}{7cm}

                $\begin{array}{l}
                    \\[-0.4cm]
                    \textit{Note: }
                    \text{For $\TxCC=\RxCC=0.9$, the ergodic capacity $\ErC$ is}
                    \\[-0.15cm]
                    \text{$7.36$ bits/symbol at $\snr=15$ dB.}
                \end{array}$
            \end{minipage}
        }

    \end{tabular}
\end{center}
\end{table}

\clearpage

\clearpage

\clearpage

\clearpage

\clearpage

\end{document}